\documentclass[a4paper,12pt]{JHEP3}

\usepackage{latexsym}
\usepackage{amsmath}
\usepackage{amsfonts}
\usepackage{mathrsfs}
\usepackage{amssymb}
\usepackage{dsfont}
\usepackage{youngtab}

\newcommand{\hal}{{\textstyle\frac{1}{2}}}

\newcommand{\ca}[1]{{\cal{#1}}}

\newcommand{\unit}{\mathds{1}}  
\newcommand{\fr}[2]{{\textstyle{\frac{#1}{#2}}}}

\newcommand{\del}{\partial}
\newcommand{\vep}{\varepsilon}
\newcommand{\csep}{\setlength\arraycolsep{2pt}}

\newcommand{\F}[4]{f^{#1#2}_{\ #3#4}}
\newcommand{\ww}{\tilde}
\newcommand{\bca}[1]{\bar{\cal{#1}}}

\newcommand{\be}{\begin{equation}}
\newcommand{\ee}{\end{equation}}
\newcommand{\ben}{\begin{displaymath}}
\newcommand{\een}{\end{displaymath}}

\newcommand{\bea}{\begin{eqnarray}}
\newcommand{\eea}{\end{eqnarray}}

\newcommand{\ft}[2]{{\textstyle {\frac{#1}{#2}} }}

\makeatletter \@addtoreset{equation}{section} \makeatother

\author{Henning Samtleben and Robert Wimmer\\
Universit\'e de Lyon, Laboratoire de Physique, UMR 5672, CNRS, \\
\'Ecole Normale Sup\'erieure de Lyon,\\
46, all\'ee d'Italie, F-69364 Lyon cedex 07, France\\
{\tt henning.samtleben, robert.wimmer\,\, @ens-lyon.fr}
}

\title{${\cal{N}}=6$ Superspace Constraints, SUSY Enhancement and Monopole Operators}

\abstract{We present a systematic analysis of the ${\cal{N}}=6$ superspace constraints 
in three space-time dimensions.
The general coupling between vector and scalar supermultiplets is encoded
in an $SU(4)$ tensor $W^i{}_j$  which is a function of the matter
fields 
and subject to a set of algebraic
and super-differential relations. We give a genuine ${\cal{N}}=6$
classification for superconformal models with polynomial interactions
and find the known ABJM and ABJ models.

We further study the issue of supersymmetry enhancement to 
${\cal N}=8$ and the role of monopole operators in this scenario.
To this end we assume the existence of a composite monopole operator
superfield which we use to
formulate the additional supersymmetries as internal
symmetries of the  ${\cal{N}}=6$ superspace constraints. From the
invariance conditions of these constraints
we derive a system of superspace constraints for the 
proposed monopole operator superfield. This constraint system defines
the composite
monopole operator superfield  analogously to the original
${\cal{N}}=6$ superspace constraints defining the dynamics of the 
elementary fields.}


\preprint{$\,$}

\begin{document}   

\section{Introduction}

The construction of three-dimensional $\ca{N}=8$ superconformal field theories of 
Bagger-Lambert-Gustavsson (BLG)
\cite{Bagger:2007jr,Gustavsson:2007vu} has triggered an immense body of work
on its relation to the dynamics of multiple M2-branes.
While manifest $\ca{N}=8$ supersymmetry singles out a unique theory 
with gauge group $SO(4)$, it has been proposed in \cite{Aharony:2008ug} that 
$N$ M2-branes located on a $\mathbb{C}^4/\mathbb{Z}_k$ singularity are described 
by a matter coupled $U(N)\times U(N)$ Chern-Simons theory of level $\pm k$ 
with manifest $\ca{N}=6$ supersymmetry, the ABJM model. For level $k=1, 2$  
the expected enhancement to $\ca{N}=8$ supersymmetry is proposed to rely on the existence
of monopole or 't Hooft operators.

The purpose of this paper is to develop an $\ca{N}=6$ superspace approach
for the formulation and classification of three-dimensional $\ca{N}=6$ theories,
to recover the known models in this formalism and to analyze the structure of the
$\ca{N}=8$ supersymmetry enhancement of the ABJM model by monopole operators 
by deriving a system of superspace constraints for these operators.

Previous superspace formulations of the ABJM model include the off-shell 
formulations in ${\cal N}=2$ \cite{Benna:2008zy}
and harmonic ${\cal N}=3$ superspace~\cite{Buchbinder:2008vi,Ivanov:2009nr},
the ${\cal N}=6$ formulation of \cite{Cederwall:2008xu} based on pure spinors
and some results in on-shell ${\cal N}=6$ superspace announced in~\cite{Bandos:2009dt}.
In the work presented here, 
we formulate and analyze the $\ca{N}=6$ superspace constraints for 
general three-dimensional gauge theories.
The matter sector
is  described by a complex scalar superfield $\Phi^i$ transforming in the fundamental representation
of the $SU(4)\sim SO(6)$  $R$-symmetry group. The gauge sector is described by a vector superfield which is 
an $SU(4)$ singlet. 
These superfields are subject to appropriate constraints to restrict the field content and we study the 
possible couplings of the 
gauge and matter superfields. 
In close analogy to the structure of $\ca{N}=8$ superspace constraints
which we have worked out in \cite{Samtleben:2009ts}, 
the set of consistent $\ca{N}=6$ theories 
can be parametrized by a hermitean $SU(4)$ tensor $W^i{}_j$, which is a function of the matter 
superfields subject to 
the following concise $SU(4)$-projection conditions:
\begin{equation}
\label{eq:intro}
\nabla_{\alpha\,ij} W^k{}_l\;\Big|_{\bf 64} = 0\ , \qquad 
W^i{}_j\cdot\Phi^k \,\Big|_{{\bf 36}} = 0 \  ,\nonumber
\end{equation}
which will be explained in detail in the main text.
The  ${\cal{N}}=6$ superspace formulation implemented here is necessarily on-shell, so that pure 
superspace geometrical considerations of the multiplet 
structure determine the dynamics of the system in terms of superfield equations 
of motions. These can be expressed in terms of the tensor $W^k{}_l$ and its super-derivatives.

We give an explicit class of solutions to the above conditions which describe 
the superconformal model with gauge group $U(N)\times U(\widetilde{N})$.
We work out the superfield and component field equations and show that the latter reproduce the
results of~\cite{Aharony:2008ug,Hosomichi:2008jb,Aharony:2008gk,Schnabl:2008wj}.
The superspace formulation that we present here provides a setting which allows the study of possible 
generalizations of these models and  the determination of quantum corrections  (to the e.o.m.) 
through symmetry considerations and by the rigidness of the $\ca{N}=6$ superspace, circumventing perturbation theory. 

The ${\cal N}=6$ superspace formalism developed in this paper provides a suitable framework
for a closer analysis of the proposed supersymmetry enhancement in the ABJM model in terms of 
monopole operators and a more explicit description of such operators.
The basic idea is to formulate the enhanced supersymmetry as
an internal $\ca{N}=2$ supersymmetry of the $\ca{N}=6$ superspace
constraint equations rather than for the Lagrangian. The additional
susy is thus an infinitesimal symmetry of the equations of motions, a
typical situation for hidden symmetries. By starting from a general 
ansatz for the additional supersymmetry transformations in
terms of monopole operators, we analyze their compatibility with the
above superspace constraints and deduce the full set of superspace 
constraints for these operators. While we leave a detailed analysis of 
this system to future work 
let us stress that in contrast to previous approaches 
\cite{Gustavsson:2009pm}
this system does not involve any additional conditions on the 
elementary fields of the theory.

We analyze two different situations. First we study the case of a
covariantly constant monopole operator, an assumption which was also made in 
the ordinary space-time approach of \cite{Gustavsson:2009pm,Kwon:2009ar}. 
We prove that under this assumptions monopole operators exist only 
in the case that the gauge group is $U(2)\times U(2)$ and recover the 
superspace version of the result given in \cite{Kwon:2009ar}. We argue 
on generals reasons that susy enhancement based on this particular operator 
ceases to exist in the quantum theory. Next we relax the assumption 
of covariant constancy and derive a system of superspace constraints 
for the proposed monopole operator which does not impose any apparent
restrictions on the dimension of the gauge group.

Eventually, this system of superspace constraints should lead to
space-time equations of motion for the composite monopole superfield, 
in analogy to our superspace analysis for the elementary superfields.
These space time e.o.m.\ for the composite monopole superfield might then describe the 
dynamics of a theory dual to the ABJM model, in the sense of the three-dimensional 
mirror symmetry \cite{Intriligator:1996ex}. This would finally be  a nonabelian gauge theory analogon of the explicit duality relations for
 two-dimensional soliton models \cite{Coleman:1974bu,Mandelstam:1975hb}.

The paper is organized as follows. In section~2 we review our conventions 
for $\ca{N}=6$ superspace and superfields.
Section~3 is devoted to an analysis of the gauge sector of three-dimensional theories and 
presents the superfield constraints to be imposed
on the super field strength in order to properly restrict the field content and dynamics.
We introduce deformations of the free Chern-Simons constraint 
which are parametrized by a hermitean $SU(4)$ tensor $W^i{}_j$.
In section~4, we describe the free matter sector and the interactions induced by
minimal coupling to the gauge sector which are parametrized by the tensor $W^i{}_j$.
In particular, we derive the full set of consistency constraints which this tensor must satisfy.
We derive the explicit component equations of motion for general $W^i{}_j$ 
and discuss their equivalence to the superfield constraints.

In section~5 we classify the conformal solutions to these consistency constraints.
We work out the full set of field equations and show that they reproduce the
ABJ model with gauge group $U(N)\times U(\widetilde{N})$\,.
Section~6 finally addresses the issue of supersymmetry enhancement 
of the ABJM model by monopole operators.
After a detailed discussion of the general properties of such operators we start from
the general ansatz for the additional supersymmetry transformations in terms 
of the monopole operators and derive the consistency conditions that are implied
by compatibility with the superspace constraints derived in the earlier sections.
We show that for covariantly constant monopole operators these conditions
necessarily imply $N=\widetilde{N}=2$. For covariantly non-constant monopole operators and general $N$ 
we derive the full system of superspace constraints for these operators and their super-derivatives.
\\

\noindent
{\bf Note added:} While finishing this work the paper \cite{Bashkirov:2010kz} appeared on the arXiv which
analyzes in a complementary 
approach some of the questions which are addressed also in the current investigations.

\section{${\cal{N}}=6$ Superspace Setup}

We briefly introduce the basic setup and our conventions for the $\ca{N}=6$ superspace calculus. For more
explicit details see the appendix. The $R$-symmetry group of the $\ca{N}=6$ susy algebra is $SO(6)\sim SU(4)$,
where we will use the $SU(4)$ notation throughout the paper. The $\ca{N}=6$ superspace
$\mathbb{R}^{2,1|12}$ is parametrized by coordinates $(x^{\alpha\beta},\theta^{\alpha\,ij})$, 
$\alpha,\beta = 1,2$ and $i,j=1,\ldots, 4$,
where $x^{\alpha\beta}$
is a real symmetric matrix and the fermionic coordinates  $\theta^{\alpha\, ij}=\theta^{\alpha\, [ij]}$ are
in the real ${\bf 6}$ of $SU(4)$, i.e.\ they 
satisfy the reality property
\bea
(\theta^{\alpha\, ij})^* =  \ft12 \epsilon_{ijkl} \theta^{\alpha\, kl}=:\theta^\alpha_{ij}
\;.
\eea

With the fermionic derivative $\del_{\alpha\, ij}\theta^\beta_{kl}=\fr{1}{2}\epsilon_{ijkl}\delta^\beta_\alpha$ 
we define the susy covariant derivatives and the susy generators as the operators, 
\begin{equation}
\label{eq:1.1}
D_{\alpha\,ij}=\del_{\alpha\,ij} + i\theta^\beta_{ij} \del_{\alpha\beta}\ ,\qquad 
Q_{\alpha\,ij}=\del_{\alpha\,ij} - i\theta^\beta_{ij} \del_{\alpha\beta}\ ,
\end{equation}
respectively, such that $\{ D_{\alpha\,ij}, Q_{\beta\,kl}\}=0$ and 
$\{ Q_{\alpha\, ij}, Q_{\beta\, kl}\} = - \{ D_{\alpha\, ij}, D_{\beta\, kl}\} = -i\epsilon_{ijkl}\, \del_{\alpha\beta}$.

For the formulation of gauge theories we introduce connections, i.e.\ covariant derivatives, on 
the considered superspace in the following way:
\csep
\begin{eqnarray}
\label{eq:1.3}
\nabla_{\alpha\beta}&=&\del_{\alpha\beta} + \mathcal{A}_{\alpha\beta}\ \ \ \quad  \textrm{with}\quad \ \ 
                   (\ca{A}_{\alpha\beta})^\dagger=-\ca{A}_{\alpha\beta} \ , \nonumber\\
\nabla_{\alpha\,ij}&=& D_{\alpha\,ij} + \mathcal{A}_{\alpha\,ij}\ \quad  \textrm{with} \quad\ \ 
            (\ca{A}_{\alpha\,ij})^\dagger=\fr{1}{2}\epsilon^{ijkl}\ca{A}_{\alpha\,kl}  \ .  
\end{eqnarray}
The connection one-forms live in a Lie algebra $\mathfrak{g}$ and are 
expanded\footnote{We include in general an extra $i$ in such expansions
for fermionic superfields in the gauge sector so that the coefficient superfields satisfy
reality conditions corresponding to their $SU(4)$ representations. See the appendix for more details.} as  
$\ca{A}_{\alpha\beta}=\ca{A}_{\alpha\beta}^M T_M$ and  $\ca{A}_{\alpha\,ij}=\ca{A}_{\alpha\, ij}^M i T_M$
 in terms of the anti-hermitian generators $(T_M)^\dagger=-T_M$
of the yet unspecified gauge group $G$.
The connection (\ref{eq:1.3}) defines a super field strength via (anti)commutators minus torsion terms, 
whose components are
\begin{eqnarray}
  \label{eq:1.4-1}
   \ca{F}_{\alpha\beta,\gamma\delta}&=&[\nabla_{\alpha\beta},\nabla_{\gamma\delta}]\quad,\quad
   \ca{F}_{\alpha\beta,\gamma\,ij}=[\nabla_{\alpha\beta},\nabla_{\gamma\,ij}] \ ,\nonumber\\
    \ca{F}_{\alpha\,ij,\beta\,kl}&=&\{\nabla_{\alpha\,ij},\nabla_{\beta\,kl}\}-i\epsilon_{ijkl}\nabla_{\alpha\beta}
     \ .
\end{eqnarray}
A given field strength has to satisfy  the Bianchi identities, which are simply obtained from the 
super-Jacobi identities 
for the covariant derivatives:\footnote{The exponent $\pi$ in the second 
identity counts the cyclic permutations where (anti)commutators are distributed corresponding to the occurrence of
bosonic/fermionic connections} 
\begin{eqnarray}
\label{eq:BI}
\sum_{\textrm{cyclic}}[\nabla_{\alpha\,ij},\{\nabla_{\beta\,kl},\nabla_{\gamma\,mn}\}]\equiv 0\ ,&&\qquad
\sum_{\textrm{cyclic}}(-1)^{\pi}\{\nabla_{\alpha\,ij},[\nabla_{\beta\,kl},\nabla_{\gamma\delta}]\}\equiv 0\ ,
 \nonumber\\
\sum_{\textrm{cyclic}}[\nabla_{\rho\,ij},[\nabla_{\alpha\beta},\nabla_{\gamma\delta}]]\equiv 0\ ,&&
\qquad\qquad\quad\ \ 
\sum_{\textrm{cyclic}}[\nabla_{\alpha\beta},[\nabla_{\gamma\delta},\nabla_{\rho\sigma}]]\equiv 0\ .
\end{eqnarray}
These identities will lead to consistency conditions for the constraints to be imposed on the super field strength.  

In addition to super gauge fields we will need matter superfields. The 
${\cal{N}}=6$ matter component multiplet $(\phi^i, \psi_{\alpha\,i})$
consists of scalar and fermion fields in the ${\bf 4}$ and ${\bf\bar{4}}$
of $ SU(4)$, respectively. Accordingly we introduce complex bosonic and fermionic 
matter superfields $\Phi^i$, $\Psi_{\alpha\,i}$ transforming in the representation 
$R$ of the gauge group which when indicate carry an upper index from the range $a,b,c,\dots$.
The complex conjugated fields
\begin{equation}
  \label{eq:1.5}
  (\Phi^{i\,a})^*=:\bar\Phi_{i\,a}\quad ,\quad (\Psi_{\alpha\,i}^a)^*=:\bar \Psi^i_{\alpha\,a}\ ,
\end{equation}
transform in the representation $\bar{R}$  and carry  a lower gauge index, as 
indicated here. However, frequently we will not write the gauge indices explicitly.

In the following sections we will impose constraints on the superfields and investigate the 
resulting dynamics.

\section{Gauge Field Constraints}\label{gaugesector}
We closely follow the methods developed in \cite{Samtleben:2009ts} for the $\ca{N}=8$ case.\footnote{For other $\ca{N}=8$ 
superfield approaches which specifically describe the BLG model
 using Nambu-brackets and pure spinors
 see \cite{Bandos:2008df, Cederwall:2008vd}} To eliminate 
unphysical degrees of freedom one imposes (partial) flatness conditions
on the bi-spinor field strength \cite{Grimm:1977xp, Sohnius:1978wk, Witten:1978xx, Witten:1985nt}, 
which is  $\ca{F}_{\alpha\, ij, \beta\, kl}$ here. In many cases this 
corresponds to an underlying geometric structure 
of twistors and pure spinors \cite{ Witten:1978xx, Witten:1985nt,Harnad:1988rs,Howe:1991mf}.  
The bi-spinor field strength in (\ref{eq:1.4-1}) contains the representations\footnote{
Decompositions of tensor products of  representations can be computed with the program LiE 
\cite{LIE} or found in \cite{Slansky:1981yr}.}
\begin{equation}
 \label{convent}
 \ca{F}_{\alpha\,ij, \beta\,kl}\sim (({\bf{2}},{\bf{6}})\otimes ({\bf{2}},{\bf{6}}))_{\mathrm{sym}} =
  ({\bf 3},{\bf 1})\oplus ({\bf 1},{\bf 15}) \oplus({\bf 3},{\bf 20}) ,
\end{equation}
where the first entry refers to the $SO(2,1)$ and the second entry to the $SU(4)$ representation.
The $SU(4)$ representations ${\bf 6}$ carried by $\ca{F}_{\alpha\,ij, \beta\,kl}$ are the real ${\bf 6}$
and consequently also the $SU(4)$ representations appearing on the r.h.s. of (\ref{convent}) are the
real ${\bf 15}$ and ${\bf 20}$.\footnote{Reality conditions can be implemented for representations
whose Dynkin labels are self-conjugated under $(r,s,t)\rightarrow (t,s,r)$, i.e.\ if $r=t$. This is the case for
the ${\bf 15}$ and $\bf{20}$ (denoted $\bf{20'}$ in \cite{Slansky:1981yr}) which have Dynkin 
labels $(1,0,1)$ and $(0,2,0)$. In the appendix we give the 
reality conditions for the tensors appearing in the following.} 

The $({\bf 3},{\bf 1})$ part in (\ref{convent}) corresponds to a second component vector field in the 
superfield expansion of
$\ca{A}_{\alpha\,ij}$ with the same gauge-transformation as the lowest component of  $\ca{A}_{\alpha\beta}$. 
Setting this part to zero imposes the so-called ``conventional constraint'' \cite{Gates:1979wg,Gates:1983nr}
for the three-dimensional $\ca{N}=6$ case and eliminates this additional component vector field.
Putting further constraints on $\ca{F}_{\alpha\,ij, \beta\,kl}$, in contrast, will not only eliminate component fields,
but also induce equations of motion for the remaining fields. We shall analyze this in more detail now.

In analogy to \cite{Samtleben:2009ts} we allow the $({\bf 1},{\bf 15})$ in (\ref{convent}) to be non-vanishing  
but set the $ ({\bf 3},{\bf 1}) \oplus({\bf 3},{\bf 20})$ part to zero (a non-vanishing $({\bf 3},{\bf 20})$ part
may be taken into account for studying higher derivative corrections). 
Given the definition (\ref{eq:1.4-1}) this constraint writes as\footnote{For the symmetry properties of the second term on the r.h.s.\ see
(\ref{eq:a13.14}).}  
\begin{equation}
\label{defW}
  \{\nabla_{\alpha\,ij} , \nabla_{\beta\,kl} \} =
      i \left(\epsilon_{ijkl} \nabla_{\alpha\beta} 
      +  \vep_{\alpha\beta} \, \epsilon_{mij[k} W^m{}_{l]} \right) \ ,
\end{equation}
where $W^i{}_j$ is an $SU(4)$ tensor transforming in the real ${\bf 15}$ of $SU(4)$ and lives in the Lie 
algebra $\mathfrak{g}$ of 
the yet unspecified gauge group $G$.
It is therefore traceless ($W^i{}_i=0$) and satisfies the hermiticity conditions
\begin{equation}
  \label{eq:wherm}
  (W^i{}_j)^\dagger = W^j{}_i\ ,
\end{equation}
where $W^i{}_j$ is expanded in the gauge algebra  as $W^i{}_j=W^{M\,i}{}_j T_M$, see the appendix for more details.
A priori $W^i{}_j$ is an independent superfield but we will see that eventually it will be a function 
of the matter superfields (\ref{eq:1.5}), i.e.\
$W^i{}_j=W^i{}_j(\Phi^k,\Psi_{\alpha\,l})$. In this
regard we will call $W^i{}_j$ the \emph{deformation potential} since it represents a deformation 
of the constraint $\ca{F}_{\alpha\,ij, \beta\,kl}=0$ for which the resulting multiplet contains exclusively a
free Chern-Simons component gauge field, i.e.\ a flat connection on $\mathbb{R}^{2,1}$, see below. 

{\bf Bianchi Identities.} For the constraint (\ref{defW})  to be consistent $W^i{}_j$ cannot be chosen 
arbitrarily but is
itself subjected to certain conditions so that the Bianchi identities (\ref{eq:BI}) are satisfied. 
The immediate nontrivial conditions on the 
superfields are given by the first two Bianchi identities in (\ref{eq:BI}), which involve the constrained
bi-spinor field strength $\ca{F}_{\alpha\,ij, \beta\,kl}$. 

Using the constraint (\ref{defW}) the first Bianchi identity imposes the condition  
\begin{eqnarray}
\label{WF}
\epsilon_{ijkl} \ca{F}_{\alpha\beta,\gamma\, mn}
\!\!&&+~\epsilon_{mnij} \ca{F}_{\gamma\alpha,\beta\, kl}  
+\epsilon_{klmn} \ca{F}_{\beta\gamma,\alpha\, ij}   =\nonumber\\
&&\vep_{\beta\gamma} \nabla\!_{\alpha\, ij}  W^p{}_{[m}\,\epsilon_{n]klp}+
\vep_{\gamma\alpha} \nabla\!_{{\beta\, kl}}  W^p{}_{[i} \,\epsilon_{j]mnp}+
\vep_{\alpha\beta}\nabla\!_{\gamma\, mn}  W^p{}_{[k} \,\epsilon_{l]ijp}\  .
\end{eqnarray}
Decomposing l.h.s.\ and r.h.s.\ of this equation 
according to their $SU(4)$ representation content,
one deduces that solvability requires the 
representation ${\bf 64}$ to vanish within the 
the tensor product 
$\nabla_{\alpha\, ij} W^k{}_l\sim {\bf 6} \otimes {\bf 15} = {\bf 6}\oplus {\bf 10}\oplus 
   {\bf \overline{10}}\oplus {\bf 64}$.
It also  implies the existence of superfields 
$\lambda_{\alpha\, ij}=\lambda_{\alpha\, [ij]}$ in the real ${\bf{6}}$, 
i.e.\ $(\lambda_{\alpha\, ij})^\dagger=\ft12\epsilon^{ijkl}\lambda_{\alpha\, kl}$,
and $\rho_{\alpha\, ij}=\rho_{\alpha (ij)}$ in the complex ${\bf \overline{10}}$,
i.e.\ $(\rho_{\alpha\, ij})^\dagger=:\bar\rho_\alpha^{\,ij}\sim{\bf 10}$,
such that  the superderivative $\nabla_{\alpha\, ij} W^k{}_l$ satisfies the condition
\bea
\label{constraint64}
&&\nabla_{\alpha\,ij} W^k{}_l\;\Big|_{\bf 64} = 0 \nonumber\\[1ex]
 && \Longrightarrow \quad
\nabla_{\alpha\,ij} W^k{}_l  ~=~  
\delta^k_{[i}\,\lambda_{\alpha}{}^{\vphantom{k}}_{j]l}
+\ft14 \delta^k_l\,\lambda_{\alpha\, ij}
+\delta^k_{[i}\,\rho_{\alpha}{}^{\vphantom{k}}_{j]l}
-\ft12\epsilon_{ijln}\,\bar\rho_{\alpha}^{\, kn}
\ .
\eea
This constraint will play a central role in the following.
If we consider $W^i{}_j$ as a function of the matter fields of the theory,
this composite superfield must satisfy (\ref{constraint64}) in order for the system 
(\ref{defW}) to be consistent.
The Bianchi identity (\ref{WF}) then fixes the fermionic field strength  
$\ca{F}_{\alpha\beta,\gamma\, ij}$ to 
\begin{equation}
\label{FL}
     \ca{F}_{\alpha\beta,\gamma\, ij} =-\ft12\vep_{\gamma(\alpha}\lambda_{\beta)\,ij}\ .
\end{equation}

Before investigating the residual Bianchi identities let us assume that we have picked a 
$W^i{}_j(\Phi^k,\Psi_{\alpha\,l})$ satisfying (\ref{constraint64}).
The integrability condition of (\ref{constraint64}) is then identically fulfilled and determines 
the superderivatives of the composite fields $\lambda_{\alpha\, ij}$ and $\rho_{\alpha\, ij}$,
\csep
\begin{eqnarray}
  \label{Dlambdarho}
\nabla_{\alpha\,ij}\lambda_{\beta\,kl} &=&
i \epsilon_{ijkl} \ca{F}_{\alpha\beta}
+2i\nabla_{\alpha\beta} W^m{}_{[k} \epsilon_{l]ijm}
+2i\vep_{\alpha\beta}V^m{}_{[k} \epsilon_{l]ijm} \ ,
\nonumber\\[1.5ex]
\nabla_{\alpha\,ij}\rho_{\beta\,kl} &=&
-i \nabla_{\alpha\beta} W^m{}_{(k} \epsilon_{l)ijm}
+i\vep_{\alpha\beta}V^m{}_{(k} \epsilon_{l)ijm}
+U_{\alpha\beta}{}^m{}_{(k} \epsilon_{l)ijm}
\nonumber\\[.5ex]
&&{}
+i\,\vep_{\alpha\beta}\left(
\ft12\epsilon_{ijmn}\,[W^m{}_k,W^n{}_l]
+\ft{1}2 \epsilon_{mij(k}\,[W^n{}_{l)},W^m{}_n] \right)\ ,
\end{eqnarray}
up to the space-time vectors and scalar  ${\cal{F}}_{\alpha\beta}$, $U_{\alpha\beta}{}^i{}_{j}$ and $V^i{}_{j}$.
The vector $ (\ca{F}_{\alpha\beta})^\dagger=- \ca{F}_{\alpha\beta}$ is an anti-hermitian $SU(4)$ 
singlet, whereas the other two tensors transform in the real ${\bf 15}$ of $SU(4)$ thus satisfying 
a hermiticity condition as given in (\ref{eq:wherm}). 
Comparing the equations (\ref{Dlambdarho}) with the second Bianchi identity 
\begin{equation}
\label{bi2}
\nabla_{\alpha\,ij}\ca{F}_{\gamma\delta,\beta\,kl} +  \nabla_{\beta\,kl}\ca{F}_{\gamma\delta,\alpha\,ij}=
    i(\epsilon_{ijkl}\ca{F}_{\gamma\delta,\alpha\beta}
      +\vep_{\alpha\beta}\nabla_{\gamma\delta}W^m{}_{[k}\,\epsilon_{l]ijm})\ ,
\end{equation}
and using (\ref{FL}), shows that it is identically fulfilled upon setting the vector
$\ca{F}_{\alpha\beta}=\vep^{\gamma\delta} \ca{F}_{\alpha\gamma, \beta\delta}$, which 
thus equals the dual bosonic field strength. With this identification one obtains that the 
integrability condition of the first equation in (\ref{Dlambdarho}), besides determining 
$\nabla_{\alpha\,ij}V^k{}_l$,  equals the third Bianchi identity in (\ref{eq:BI}). 
 
{\bf Super Chern-Simons e.o.m.} 
With a chosen deformation potential $W^i{}_j(\Phi^k,\Psi_{\alpha\,l})$  the derived superfields 
$\lambda_{\alpha\,ij}$, $\rho_{\alpha\,ij}$, etc. are given functions of the matter superfields. In particular, 
the first equation of (\ref{Dlambdarho}) together with (\ref{constraint64}) gives 
the following Chern-Simons equations of motion for the bosonic field strength:
\begin{equation}
\label{eomCS}
 \ca{F}_{\alpha\beta}=\fr{1}{24\, i}\, \epsilon^{ijkl}\nabla_{ij\,(\alpha}\lambda_{\beta)\,kl}= 
       \fr{1}{15\, i}\, \nabla^{ik}_{(\alpha}\nabla^{\vphantom{i}}_{\beta)\,jk} W^j{}_i  ,
\end{equation}
which for $W^i{}_j = 0$ reduces to the free CS-e.o.m.\
A priori, with (\ref{eomCS}) the fourth Bianchi identity in (\ref{eq:BI}), which takes the form
\begin{equation}
\label{eq:2.11}
\nabla^{\alpha\beta}\ca{F}_{\alpha\beta}=0\ ,
\end{equation}
might give rise to yet another condition. Evaluating the l.h.s.\ of (\ref{eq:2.11}) with $\ca{F}_{\alpha\beta}$
given by (\ref{eomCS}) shows that this Bianchi identity is also identically satisfied without 
any further conditions and thus (\ref{constraint64}) automatically defines a covariantly conserved current, the r.h.s.\ 
of (\ref{eomCS}).
\\

Consequently, with a given choice for the deformation potential $W^i{}_j$ which satisfies the 
constraint (\ref{constraint64}) all Bianchi identities are identically fulfilled upon imposing the 
 CS equations of motion (\ref{eomCS}), the fermionic field strength is given by (\ref{FL}). 
Therefore  (\ref{constraint64}) represents the only restriction on the 
choice of the deformation potential $W^i{}_j(\Phi^k,\Psi_{\alpha\,l})$  for the gauge field constraint 
(\ref{defW}) to be consistent. We will address the issue of 
component field equations and their equivalence to the constraint (\ref{defW}) at a later point when we have 
discussed the matter sector which  couples non-trivially to the gauge sector.

\section{Matter Field Constraints}
\label{sec:msf}

The ${\cal{N}}=6$ matter multiplet $(\phi^i, \psi_{\alpha\,i})$
consists of scalar and fermion component fields in the complex ${\bf 4}$ and ${\bf\bar{4}}$
of $SU(4)$, respectively. It is therefore natural to encode this multiplet in a scalar superfield 
$\Phi^i$  in the ${\bf 4}$ to be subjected to  appropriate constraints. At first order in 
$\theta^{\alpha\,jk}$ this superfield contains a fermionic component 
$\chi^{\,i}_{\alpha\, kl}$
which decomposes into irreps as 
${{\bf 4}\otimes{\bf \bar{6}} = {\bf\bar{4}}\oplus{\bf \overline{20}}}$. The
super- and gauge-covariant way to project out the ${\bf \overline{20}}$ in 
accordance with the field content is to impose the condition
\begin{equation}
\label{dc2}
\nabla_{\alpha\,ij} \Phi^k \,\Big|_{{\bf \overline{20}}}=0\quad
\Longrightarrow\qquad
\nabla_{\alpha\,ij} \Phi^k = i\,  \delta^k_{[i}\Psi^{\vphantom{k}}_{j]\,\alpha}\ ,
\end{equation}
with a fermionic superfield $\Psi_{\alpha\,i}$ in the ${\bf \bar 4}$, which is defined by this equation.
Explicitly this gives\footnote{Here we anticipate that $\ca{A}_{\alpha\,ij}|_{\theta =0}=0$, i.e.\ we omit 
the term $\epsilon^{ijlm} \ca{A}_{\alpha\,lm}\cdot\Phi^k|_{\theta =0}$. 
This will be justified below.} 
$\chi^{\,k}_{\alpha\, ij}= i\, \delta^k_{[i}\Psi^{\vphantom{k}}_{j]\,\alpha}|_{\theta=0}
=:i\, \delta^k_{[i}\psi^{\vphantom{k}}_{j]\,\alpha}$\,.

In the following we derive the consequences implied by combining (\ref{dc2}) 
with the vector superfield constraint (\ref{defW}). 
Using the latter, the integrability condition of (\ref{dc2}) takes the form
\begin{equation}
\label{dic}
\epsilon_{ijkl}  \nabla_{\alpha\beta}\Phi^m  
+ \vep_{\alpha\beta} \,\epsilon_{nij[k} W^n{}_{l]} \cdot \Phi^m
= 
- \nabla_{\alpha\,ij} \Psi^{\vphantom{n}}_{\beta\,[k} \delta^m_{l]}
- \nabla_{\beta\,kl} \Psi^{\vphantom{n}}_{\alpha\,[i} \delta^m_{j]}
\;,
\end{equation}
where $W^i{}_{j} \cdot \Phi^k$ denotes the action
of the algebra valued $W^i{}_{j}$ onto the scalar superfield.
This poses an algebraic constraint on the deformation potential $W^i{}_j$ since the 
unpaired ${\bf 36}$ in
\begin{equation}
  \label{eq:wf}
  W^i{}_{j}\cdot\Phi^k\sim {\bf 15}\otimes{\bf 4}={\bf 4}\oplus{\bf 20}\oplus{\bf 36}
\end{equation}
does not drop out of equation (\ref{dic}). Explicitly this constraint writes as 
\setlength\arraycolsep{0pt}
\begin{eqnarray}
\label{ac36}
W^i{}_j\cdot\Phi^k \,\Big|_{{\bf 36}} = 0 \qquad \Longleftrightarrow \qquad
W^{(i}{}_j\cdot\Phi^{k)} = 
\ft1{5} \, \delta^{(i}_j\, W^{k)}_{\vphantom{i}}{}_n\cdot\Phi^n \ ,
\end{eqnarray}
and thus the deformation potential has to be a function of the matter fields, as mentioned before. 
In addition to the super-differential constraint (\ref{constraint64})  
this algebraic constraint will be the main restriction on the possible 
choices for the deformation potential
$W^i{}_j(\Phi^k,\Psi_{\alpha\,l})$, which fixes the details of the dynamics. In the following we will
refer to these two constraints (\ref{constraint64}) and  (\ref{ac36}), which determine the set of possible 
models, as the \emph{$W$-constraints}, as in the $\ca{N}=8$ case \cite{Samtleben:2009ts}.  

With the condition (\ref{ac36})  the integrability condition (\ref{dic}) can be resolved to 
determine the superderivative $\nabla_{\alpha\,ij} \Psi_{\beta\,k}$ :
\bea
\nabla_{\alpha\,ij} \Psi_{\beta\,k} 
&~=~& 
-2\epsilon_{ijkl} \nabla_{\alpha\beta} \Phi^l
+ 
\ft12\vep_{\alpha\beta} \left(
\epsilon_{ijmn}\,W^m{}_{k}\cdot \Phi^n
+
\ft3{5}\, \epsilon_{ijkl}\,W^l{}_{n}\cdot \Phi^n \right)
\;,
\label{DPsi}
\eea

The analogous equations of this subsection for the complex conjugated fields $(\Phi^i)^*=:\bar\Phi_i$ and 
$(\Psi_{\alpha\,i})^*=:\bar\Psi^i_{\alpha}$ are obtained by complex conjugation. The necessary relations and conventions 
are given in the appendix. 

{\bf Superfield e.o.m.} The integrability conditions of (\ref{DPsi}) together with the 
gauge field constraint (\ref{defW}) 
and the various constraint relations and Bianchi identities
yield the fermion superfield e.o.m.\ for $\Psi_{\alpha\,i}$,
\begin{equation}
  \vep^{\beta\gamma}\nabla_{\alpha\beta}\Psi_{\gamma\,i} =
                  \ft{i}2\, \lambda_{\alpha ij}\cdot \Phi^j
              - \fr{i}{10}\, \rho_{\alpha ij}\cdot \Phi^j 
                  -\ft15\, W^j{}_i \cdot\Psi_{\alpha\,j} \  .
  \label{eomF}
\end{equation}
Its superderivative gives the bosonic superfield e.o.m.\ for $\Phi^i$:
\bea
\nabla^2 \Phi^i &~=~&
                    \ft14\,\vep^{\alpha\beta} \left(  \lambda^{ij}_{\alpha}\cdot\Psi_{\beta\, j}
 +\ft1{5}\, \bar\rho^{ij}_{\alpha}\cdot\Psi_{\beta\, j}\right) 
  \nonumber\\[.5ex]
                 &&{} +\ft25\, V^i{}_j \cdot\Phi^j 
                         +\ft1{25}\, W^i{}_j\cdot(W^j{}_k\cdot\Phi^k) 
                         -\ft1{20}\, W^j{}_k\cdot(W^k{}_j\cdot\Phi^i) \  .
                         \label{eomB}
\end{eqnarray}
The check that no other integrability relations descend from (\ref{eomF})
requires the double superderivative of the algebraic constraint (\ref{ac36}).
By virtue of the same constraint 
one can recast  the scalar self-interaction involving $W^i{}_j $ in different forms. Equations 
(\ref{eomF}), (\ref{eomB}) together with the Chern-Simons equation (\ref{eomCS})
for the field strength constitute the complete set of superfield e.o.m.

\subsubsection*{ Superfield expansion and equivalence to component e.o.m.} 

We briefly discuss the component content of the so far developed superspace expressions. 
This was  explained in detail in \cite{Samtleben:2009ts} for the $\ca{N}=8$ case and we can be rather short here and refer
to  \cite{Samtleben:2009ts} for more details. The very same structure and methods as in \cite{Samtleben:2009ts}
apply mutatis mutandis to the $\ca{N}=6$ case considered here. 

An essential ingredient in the derivation of component expressions and the proof of the equivalence 
between the on-shell component multiplet and the superspace constraints is the so called 
``transverse gauge''. It allows one to formulate recursion relations for the superfield expansion 
of various expressions, explicitly
\begin{equation}
  \label{eq:transgauge}
  \theta^{\alpha\, ij}\ca{A}_{\alpha\, ij}=0 \quad \Rightarrow \quad 
       \ca{R}:= \theta^{\alpha\, ij}\nabla_{\alpha\, ij}=\theta^{\alpha\, ij}\del_{\alpha\, ij}\ ,
\end{equation}
so that the recursion operator $\ca{R}$ satisfies 
$\ca{R}(\theta^{\alpha_1 i_1 j_1}\ldots \theta^{\alpha_n  i_n j_n})=
n\ \theta^{\alpha_1  i_1 j_1}\ldots \theta^{\alpha_n  i_n j_n}$.

This method was developed in \cite{Harnad:1984vk,Harnad:1985bc} for super Yang-Mills theories, where all 
superfields have a geometric origin in a (higher dimensional) vector superfield. The  cases   considered 
here and in   
\cite{Samtleben:2009ts} lead to modifications due to that fact that the matter superfields 
are not in this sense geometrically related to the vector superfields and especially 
due to the constraints on the deformation potential $W^i{}_j$, which is 
otherwise not specified at this point.  

Contracting the constraints (\ref{dc2}), (\ref{DPsi}), (\ref{defW}) with $\theta^{\alpha\, ij}$  and the 
Bianchi identity (\ref{FL}) with  
$\theta^{\gamma\, ij}$  one obtains the recursion relations
\csep
\begin{eqnarray}
\label{eq:drec}
\ca{R}\,\Phi^i&=&i \theta^{\alpha ij}\Psi_{\alpha\,j}\ \ ,\nonumber\\
\ca{R}\,\Psi_{\beta\,k}&=&-4\theta^\alpha_{kl}\nabla_{\alpha\beta}\Phi^l
                  +(\theta_{\beta\,mn}W^m{}_k\cdot\Phi^n+\fr{3}{5}\theta_{\beta\,kl}W^l{}_n\cdot\Phi^n)   \ \ ,\nonumber\\
(1+\ca{R})\,\ca{A}_{\beta\,kl}&=&2i(\theta^\alpha_{kl}\ca{A}_{\alpha\beta}+\theta_{\alpha\,m[k}W^m{}_{l]})   \ \ , 
\nonumber\\
\ca{R}\,\ca{A}_{\alpha\beta}&=& \fr{1}{2}\theta^{\gamma\,ij}\vep_{\gamma(\alpha}\lambda_{\beta)ij}  \ \ .
\end{eqnarray}
These recursion relations define the order $n+1$ in $\theta$ of the l.h.s.\ expressions in terms of the order $n$ in $\theta$ 
of the expressions on the r.h.s.\ Therefore all superfields are expended in terms of the lowest 
component fields\footnote{The composite fields $W^i{}_j$ and $\lambda_{\alpha\,ij}$ are given 
functions of the matter fields $\Phi^i$, $\Psi_{\alpha\,i}$ and thus their lowest components are functions 
of the here given component multiplet.}
\begin{equation}
  \label{eq:sfex}
  \phi^i:=\Phi^i|_{\theta=0}\ \ ,\ \ \psi_{\alpha\,i}:=\Psi_{\alpha\,i}|_{\theta=0}\ \,\ \ 
     A_{\alpha\beta}:=\ca{A}_{\alpha\beta}|_{\theta=0}\ ,
\end{equation}
which represent the $\ca{N}=6$ CS-matter component multiplet (note  that (\ref{eq:drec}) implies that 
$\ca{A}_{\alpha\,ij}$ has no component at $\theta=0$). Consequently, the lowest component of the superfield e.o.m.\
(\ref{eomCS}) and (\ref{eomF}), (\ref{eomB}) gives automatically the e.o.m.\ for these component fields. 

It is a general feature
of this approach, that many equations can be considered as component field expressions but 
are at the same time
superfield expressions and thus automatically susy covariant. There is no need to check
supersymmetry via component field susy transformations. Nevertheless, the component susy transformations 
are easily obtained by acting with the susy operators $Q_{\alpha\,ij}$ (\ref{eq:1.1}) 
on the superfields expanded to order $\ca{O}(\theta)$  
(modulo restoring supergauge transformations, see \cite{Samtleben:2009ts} for details). As a matter of fact 
the situation is even simpler. The recursions (\ref{eq:drec}) actually resemble the component susy 
transformations. 
By replacing 
 in (\ref{eq:drec}) $\ca{R}\,\Phi^i$, $\ca{R}\,\Psi_{\alpha\,i}$, $\ca{R}\,\ca{A}_{\alpha\beta}$ with 
the susy transformations of the component fields, 
i.e.\ $\delta\phi^i$, $\delta\psi_{\alpha\,i}$, $\delta A_{\alpha\beta}$, and 
all fermionic coordinates $\theta^{\alpha\,ij}$ on the r.h.s.\ with susy-parameters $\epsilon^{\alpha\,ij}$
one directly obtains the susy transformations for the component multiplet (\ref{eq:sfex}).

This shows that the basic constraints (\ref{dc2}), (\ref{defW}) imply the on-shell component
 multiplet (\ref{eq:sfex}) 
with respective
susy transformations. To prove that these two descriptions are actually equivalent, i.e.\ that the 
superspace constraints
do not imply any further conditions, one has to show that the on-shell component multiplet implies 
superfields satisfying the 
given constraints. We will not outline this here but refer the reader to  \cite{Samtleben:2009ts} where 
the proof was given  
for the $\ca{N}=8$ case, to convince oneself along the same lines 
that the same is true for the here considered $\ca{N}=6$ case.

\section{Conformal Gauge Theories, ABJM}\label{sec3}

We start by reviewing the structure of superspace constraints 
identified so far. The matter sector of these three-dimensional gauge theories 
is described by a scalar superfield subject to the constraint (\ref{dc2})
\bea
\nabla_{\alpha\,ij} \Phi^k \,\Big|_{{\bf \overline{20}}} &=& 0  \;.
\label{conP}
\eea
The full theory is then identified by specifying their gauge algebra $\mathfrak{g}$
and by choosing $W^i{}_j(\Phi, \Psi)$ in (\ref{defW}) as a 
function of the matter superfields of the theory.
This choice of the deformation potential $W^i{}_j$ is not arbitrary but 
must satisfy two independent superfield conditions, the $W$-constraints 
(\ref{constraint64}) and (\ref{ac36}):
\bea
\nabla_{\alpha\,ij} W^k{}_l\;\Big|_{\bf 64}  &=& 0\;,
\label{conW1}\\[1ex]
W^i{}_j\cdot\Phi^k \,\Big|_{{\bf 36}} &=& 0
\;.
\label{conW2}
\eea
The first equation requires that the deformation potential $W^i{}_j$ depends on the matter fields
in such a way that (\ref{conW1}) is satisfied as a consequence of (\ref{conP}).
In contrast, equation (\ref{conW2}) also explicitly contains the action of the
gauge group on the matter fields and will thus put further restrictions on
the possible gauge groups.

\subsection{Conformal Deformation Potentials}

In the following we explicitly indicate the gauge group structure. 
The matter superfields $\Phi^i$, $\Psi_{\alpha\,i}$  are taken to transform in some representation 
$R$ (which we denote by an upper index from the range 
$a,b,\ldots$)  of the to be specified gauge group $G$. They are thus denoted by 
$\Phi^{i\,a}$, $\Psi_{\alpha\,i}^a$, whereas the superfields of the gauge sector, which are in the
adjoint representation, are represented as matrices $\ca{A}_{\alpha\beta}{}^a{}_b$ etc.\ and act 
by matrix multiplication on the matter fields. The complex conjugated matter fields transform 
in the conjugate representation $\bar R$ and thus carry a lower index, see (\ref{eq:a13.10}). 

The constraint (\ref{defW}) implies that the composite field $W^i{}_j$ has canonical dimension one. Given that 
the scalar fields in three dimensions have canonical dimension $\frac{1}{2}$, scale invariance implies that 
with a polynomial ansatz  $W^i{}_j$ is
bilinear in the scalar superfields\footnote{This dimensional analysis excludes a possible dependence
on the fermionic superfields $\Psi_{\alpha\,i}$, which have dimension one, for a polynomial ansatz. Higher order 
corrections with non-polynomial deformation potentials allow for more possibilities.} $\Phi^{i\,a}$. 
The most general ansatz in the ${\bf 15}$ of $SU(4)$ is
given by
\begin{equation}
  \label{Wconf}
  (W^i{}_j)^a{}_b := \F{a}{c}{b}{d}\,
          \left(\Phi^{i\,d} \bar\Phi_{j\,c}
                       -\ft14 \delta^i_j\,\Phi^{k\,d} \bar\Phi_{k\,c} \right)
\end{equation}
with dimensionless constants $\F{a}{c}{b}{d}$.  The potential $  (W^i{}_j)^a{}_b$  is supposed to be the matrix 
that acts on 
fields in the representation $R$. The hermiticity condition (\ref{eq:wherm}) requires
that $(\F{a}{c}{b}{d})^*=\F{b}{d}{a}{c}$.  
By construction  $W^i{}_j$ has to be an element of the Lie algebra $\mathfrak{g}$  of $G$,
therefore gauge covariance requires that $\F{a}{c}{b}{d}$ is an invariant tensor of the
gauge group. Before specifying further the allowed gauge groups $G$ and representations
$R$ we check the the $W$-constraints  (\ref{Wconf}), (\ref{conW1}). 

It is straightforward to see that (\ref{Wconf}) is a solution to (\ref{conW1})
as a consequence of (\ref{conP}): since  $\nabla_{\alpha\,ij}W^k{}_l$ is composed of
a single $\Phi^i$ and a single $\bar\Psi^j_{\alpha}$ (or their complex conjugates) it transforms 
in the tensor product 
${\bf 4}\otimes {\bf {4}}={\bf 6}\oplus {\bf 10}$ and the c.c. thereof, which does not 
contain a ${\bf 64}$\,.
To solve the algebraic constraint (\ref{conW2}) we evaluate the action of (\ref{Wconf})
on a scalar field and extract the contribution ${\bf 36}$ (\ref{ac36}):
\begin{equation}
  \label{36con}
  (W^i{}_j\cdot\Phi^k)^a\big |_{{\bf 36}}=\F{a}{c}{b}{d}
                \left(\delta^l_j\Phi^{i(b}\Phi^{d)k}-\fr{2}{5}\delta^{(i}_j\Phi^{k)(b}\Phi^{d)l} \right)\bar\Phi_{l\,c}\ .
\end{equation}
This shows that the tensor $\F{a}{c}{b}{d}$ has to be antisymmetric
in its indices $[bd]$ and thus by complex conjugation  also in its
upper indices $[ac]$, i.e.\ 
\begin{equation}
  \label{ften}
  \F{a}{c}{b}{d}=\F{[a}{c]}{[b}{d]}\ .
\end{equation}

Interestingly, this solution to the $W$-constraints occupies all of the 
allowed $SU(4)$ representation content, i.e.\  $W^i{}_j\cdot\Phi^k\sim {\bf 4}\oplus{\bf 20}$. This 
differs from the conformal solution in the $\ca{N}=8$ case, where $W\cdot\Phi$ has no 
component in the allowed ${\bf 8}$ of $SO(8)$ \cite{Samtleben:2009ts}.

What are the restrictions on the gauge group $G$ and the representation $R$\,? Since 
$W^i{}_j\in\mathfrak{g}$, gauge covariance translates into the 
quadratic condition~\cite{Bergshoeff:2008bh}
\bea
(\F{a}{c}{b}{d}\, \F{d}{g}{e}{h} - \F{g}{c}{h}{d}\,\F{d}{a}{e}{b})+
(\F{c}{a}{e}{d}\,\F{d}{g}{b}{h}-\F{g}{a}{h}{d}\,\F{d}{c}{b}{e})=0
\;,
\label{quadf}
\eea
which can be obtained by comparing  $W^i{}_j\cdot W^k{}_l$ evaluated as commutator and
by the action of $W^i{}_j$ on the scalars in $W^k{}_l$, respectively. This condition is 
identical with the \emph{fundamental identity}
of (hermitean) three-algebras \cite{Bagger:2008se, Akerblom:2009gx} and shows how our result is related to this structure. 
We will not elaborate 
on this point, but rather make a more definite statement: Writing the Lie algebra valued deformation
potential as $(W^i{}_j)^a{}_b=W^{M\,i}{}_j(T_M)^a{}_b$ and using the hermiticity condition for 
$\F{a}{c}{b}{d}$ and that it is an invariant tensor one obtains
\begin{equation}
  \label{fsol}
  \F{a}{c}{b}{d}\ \rightarrow\  \overset{L}{\underset{l=1}{\bigoplus}} 
                 \ \frac{1}{g_{(l)}}\ \kappa^{MN}_{(l)}T^{(l)}_M{}^a{}_b \ T^{(l)}_N{}^c{}_d\ ,
\end{equation}
where we assumed a product $G=G_1\times\ldots\times G_L$ for the gauge group. The quantities 
$ \kappa_{(l)}^{MN}$ are the inverse of 
$\kappa^{(l)}_{MN}=\mathrm{Tr}_R(T^{(l)}_MT^{(l)}_N)$ for each factor in the gauge group and 
$g_{(l)}$ are so far arbitrary constants. 

This very same expression with the antisymmetry conditions given in (\ref{ften}) was classified in a beautiful 
analysis in 
\cite{Schnabl:2008wj}. The allowed groups are $U(N)\times U(\widetilde{N})$, $SU(N)\times SU(N)$ with matter fields
in the bifundamental representation and $g_{(1)}=-g_{(2)}=:g$, 
as well as  $U(N)$, $Sp(N)$ with matter in the fundamental representation. most of the time we will 
express our results in a more compact notation using the symbol $\F{a}{c}{b}{d}$. Of 
particular interest will be the  $U(N)\times U(\widetilde{N})$ case for which we give here the 
solution \cite{Schnabl:2008wj} and we also
indicate how
to translate the compact notation into the explicit  $U(N)\times U(\widetilde{N})$ notation. 
Every gauge index  encountered  so far
becomes a double index of the bifundamental representation 
$R=(\mathbf{N},\overline{\mathbf{\widetilde{N}}})$ as follows:
\begin{align}
\label{schnabel}
 \Phi^{i\,a} \ \ \rightarrow{}&\ \  \Phi^{i\,a}{}_{\tilde a}\ \ ,\ \ 
               \bar \Phi^i{}_b\  \rightarrow \ \bar\Phi^{i\,\tilde b}{}_{b}, \ \ \textrm{etc.}\ ,\nonumber\\[12pt]
 \hspace{-5mm} \F{a}{c}{b}{d} \  \rightarrow{}& \  
      \fr{1}{g}
        (\delta^a_d\,\delta^c_b-\fr{1}{N}\delta^a_b\,\delta^c_d\,)\, \delta^{\ww b}_{\ww a}\,\delta^{\ww d}_{\ww c}
           -\fr{1}{g}\, \delta^a_b\,\delta^c_d\,(\delta^{\ww b}_{\ww c}\,\delta^{\ww d}_{\ww a}
                     -\fr{1}{\widetilde{N}}\delta^{\ww b}_{\ww a}\,\delta^{\ww d}_{\ww c}\,) \nonumber \\[5pt]
    {}&\hspace{5mm} +\fr{1}{g}(\fr{1}{N}-\fr{1}{\widetilde{N}})\,
                       \delta^a_b\,\delta^c_d\,\delta^{\ww b}_{\ww a}\,\delta^{\ww d}_{\ww c}
       \ \ =\ \   \fr{1}{g}(\delta^a_d\,\delta^c_b\, \delta^{\ww b}_{\ww a}\,\delta^{\ww d}_{\ww c}
                                      -  \delta^a_b\,\delta^c_d\,\delta^{\ww b}_{\ww c}\,\delta^{\ww d}_{\ww a}\,)  \ .
\end{align}

The first expression for $ \F{a}{c}{b}{d}$ gives the explicit decomposition according to the contributions of 
the different group factors. 
The first two terms are  $su(N)\otimes\unit$ and  $\unit\otimes su(\widetilde{N})$ whereas the third term
 depicts the respective 
$u(1)$ contributions. As one can see, for $\widetilde{N}=N$ the $u(1)$ factors cancel 
and $SU(N)\times SU(N)$ is contained as
a special case in (\ref{schnabel}) and cannot be distinguished from $U(N)\times U(N)$ on the basis of the 
tensor $ \F{a}{c}{b}{d}$. Also the case $U(N)$ is contained as a special case by setting the range for the tilded 
indices $\widetilde{N}=1$. In the last equality we collect the terms in a more compact form.

We have thus classified all conformal $\ca{N}=6$ gauge theories. We want to emphasize that this 
classification is 
more complete than previous ones in the sense that we did not have to assume that the 
theory descends from a particular $\ca{N}=2$ theory
 \cite{Aharony:2008ug}. This is a genuine  $\ca{N}=6$ classification. Our result is 
in accordance with a complementary approach \cite{Huang:2010rn}, where possible scattering amplitudes
for conformal $\ca{N}=6$ theories where studied, also without any reference to a particular 
theory with less susy. We would also like to mention that parts of the  
structure developed here were already discussed in \cite{Bandos:2009dt}.  

We will now evaluate the
general expressions derived in the previous sections
for the conformal deformation potential  (\ref{Wconf}).

\subsection{Superfield Equations and Lagrangian}

To obtain the explicit form of the superfield e.o.m.\ one first needs 
the composite fields appearing in (\ref{eomF}), (\ref{eomB}) and (\ref{eomCS})
for the deformation potential (\ref{Wconf}). They 
can be extracted from (\ref{constraint64}) and the first equation in (\ref{Dlambdarho}):
\begin{equation}
  \label{compf}
  \lambda_{\alpha\,ij}=\fr{4}{5}\,\nabla_{\alpha\,k[i}W^k{}_{j]}\  , \ \ 
  \rho_{\alpha\,ij}=\fr{2}{3}\,\nabla_{\alpha\,k(i}W^k{}_{j)}\  ,\ \
  V^i{}_j=-\fr{i}{8}\epsilon^{iklm}\nabla_{\alpha\,kl}\lambda^\alpha_{mj}\ .
\end{equation}
The explicit expressions, evaluated for (\ref{Wconf}) are given in the appendix (\ref{compconf}). Inserting these in
 (\ref{eomF}), (\ref{eomB}) and (\ref{eomCS}) gives the superfield e.o.m.\ for the   
superconformal theories classified above. For the Chern-Simons (\ref{eomCS}), the fermionic (\ref{eomF}) and 
the bosonic (\ref{eomB}) equations of motions one obtains:
\begin{align}
  \label{csconf}
  \ca{F}_{\alpha\beta}{}^a{}_b={}&\fr{1}{2}\ \F{a}{c}{b}{d}\left(
        \Phi^{k\,d}\nabla_{\alpha\beta}\bar\Phi_{k\,c}-\nabla_{\alpha\beta}  \Phi^{k\,d}\bar\Phi_{k\,c}+
        \fr{i}{2}\Psi^d_{k\,(\alpha}\bar\Psi^k_{\beta)\,c}\right) \ ,\\[12pt]
\label{confeom}
   \hspace{-6.5pt}\vep^{\beta\gamma}\nabla_{\alpha\beta}\Psi^a_{\gamma\,i}={}& \fr{1}{4}\F{a}{c}{b}{d} \left(
    \Psi^b_{\alpha\,i}(\Phi\bar\Phi)^d{}_c-2(\Psi_{\alpha}\Phi)^{bd}\,\bar\Phi_{i\,c}
       -\epsilon_{ijkl}\Phi^{j\,b}\Phi^{k\,d}\bar\Psi_{\alpha\,c}^l \right)\ , \nonumber\\[12pt]
 \nabla^2 \Phi^{i\,a} ={}& \ca{V}^{i\,a}_{\mathrm{bos}}\, +\nonumber\\ 
           {}& \fr{i}{4}\, \vep^{\alpha\gamma} \F{a}{c}{b}{d}\left(
          (\Phi\Psi_{\alpha})^{bd}\,\bar\Psi^i_{\gamma\,c}
         -\hal \Phi^{i\,b}(\Psi_{\alpha}\bar\Psi_{\gamma})^d{}_c+
       \hal\epsilon^{ijkl}\bar\Phi_{j\,c}\Psi^b_{\alpha\,k}\Psi^d_{\gamma\,l} \right),                 
\end{align}
where $(\Phi\bar\Phi)^d{}_c=\Phi^{d\,k}\bar\Phi_{k\,c}$ etc.\ is short for contracted $SU(4)$ indices.
The dual field strength $ \ca{F}_{\alpha\beta}{}^a{}_b$ is the matrix in the representation $R$ acting on fields  $\Phi^{i\,a}$, 
for example. Under the identification (\ref{schnabel}) with
$R=(\mathbf{N},\overline{\mathbf{\widetilde{N}}})$ 
it thus decomposes as
\begin{equation}
  \label{FUU}
  \ca{F}_{\alpha\beta}{}^a{}_b \rightarrow  
\ca{F}_{\alpha\beta}{}^a{}_b\, \delta^{\ww b}{}_{\ww a}
-\delta^a{}_{b}\,\ww{\ca{F}}_{\alpha\beta}{}^{\ww a}{}_{\ww b}\ .
\end{equation}
and the covariant derivative, also acting in the bifundamental representation, is given by
\begin{equation}
  \label{dbifu}
  \nabla_{\alpha\beta}\Phi^{i\,a}{}_{\ww a}=\del_{\alpha\beta}\Phi^{i\,a}{}_{\ww a} 
     + {\cal A}_{\alpha\beta}{}^a{}_b\Phi^{i\,b}{}_{\ww a}
     - \Phi^{i\,a}{}_{\ww b}\,\ww{\cal A}_{\alpha\beta}{}^{\ww b}{}_{\ww a}\ .
\end{equation}
 
The self-interaction in the scalar field e.o.m., $\ca{V}^{i\,a}_{\mathrm{bos}}$, is a derivative of the 
bosonic potential
\begin{equation}
  \label{vprim}
  \ca{V}^{i\,a}_{\mathrm{bos}}=-\fr{1}{4}\left(
      \F{a}{c}{h}{d}\F{d}{g}{e}{b}-\hal\F{c}{a}{e}{d}\F{d}{g}{b}{h}\right)
      (\Phi\bar\Phi)^e{}_c(\Phi\bar\Phi)^h{}_g\Phi^{i\,b}\ .
\end{equation}

We can now integrate back these e.o.m.\ to obtain the Lagrangian from which they can be derived. To
make contact with the existing literature we write space-time vectors for the bosonic part in the vector notation, 
see appendix \ref{so3}. The result is:
\csep
\begin{align}
  \label{labjm}
  \ca{L}_{\mathrm{ABJM}}={}&-\mathrm{tr}\,\nabla_{\mu}\Phi^i\nabla^\mu\bar\Phi_i
         -\fr{i}{4}\mathrm{tr}\,\bar\Psi^{\alpha\,i}\nabla_{\alpha\beta}\Psi^{\alpha}_i 
           -\ca{V}_{\mathrm{bos}}+\ca{L}_{\mathrm{Yuk}}\nonumber\\[5pt]
  {}&-2\,g\, \vep^{\mu\nu\lambda}\,\mathrm{tr}
  (\ca{A}_\mu\del_\nu \ca{A}_\lambda+ \fr{2}{3}\ca{A}_\mu \ca{A}_\nu \ca{A}_\lambda
  - \ww{\ca{A}}_\mu\del_\nu\ww{ \ca{A}}_\lambda -  \fr{2}{3}\ww{\ca{A}}_\mu\ww{ \ca{A}}_\nu\ww{\ca{A}}_\lambda)\, ,
\end{align}
where  $\ca{A_\mu}$, $\ww{\ca{A}}_\mu$ are the gauge fields (in vector notation)  as given below (\ref{csconf}). The 
sextic bosonic potential writes as
\csep
\begin{align}
  \label{vbos}
  \ca{V}_{\mathrm{bos}}={}&\frac{1}{24}\left(
      \F{a}{c}{h}{d}\F{d}{g}{e}{b}-\hal\F{c}{a}{e}{d}\F{d}{g}{b}{h}\right)
      (\Phi\bar\Phi)^e{}_c(\Phi\bar\Phi)^h{}_g(\Phi\bar\Phi)^{b}{}_a\nonumber\\
      ={}&-\frac{1}{48 g^2}\mathrm{tr}\,(\Phi^i\bar\Phi_i\Phi^j\bar\Phi_j\Phi^k\bar\Phi_k +
                \bar\Phi_i\Phi^i\bar\Phi_j\Phi^j\bar\Phi_k\Phi^k \nonumber\\
   &\hspace{18mm} + 4\,\Phi^i\bar\Phi_j\Phi^k\bar\Phi_i\Phi^j\bar\Phi_k
          - 6\,\Phi^k\bar\Phi_i\Phi^i\bar\Phi_k\Phi^j\bar\Phi_j)\, , 
\end{align}
and the Yukawa interaction written in the  compact notation is
\begin{align}
  \label{yuk}
  \ca{L}_{\mathrm{Yuk}} = \fr{i}{8}\,\vep^{\alpha\gamma}\F{a}{c}{b}{d}&\left\{
  (\Phi\Psi_\alpha)^{bd}\,(\bar\Phi\bar\Psi_\gamma)_{ac} - \hal (\Phi\bar\Phi)^b{}_a (\Psi_\alpha\bar\Psi_\gamma)^d{}_c
    \right. \nonumber\\[3.5pt]
   {}&\ \ +\left.\left(\fr{1}{4}\ 
     \epsilon_{ijkl}\, \Phi^{i\,b}\Phi^{j\,d}\bar\Psi^k_{\alpha\,a}\bar\Psi^l_{\gamma\,c}-\ c.c.  \right)\right\} ,
\end{align}
where we again indicated summed $SU(4)$ indices by parentheses. In lowest order in the $\theta$ expansion
these expressions reproduce the component Lagrangians of \cite{Aharony:2008ug,Hosomichi:2008jb,Aharony:2008gk}.

Consistency of the quantum theory requires the gauge invariance of 
$\exp(iS_{CS})$ and thus determines the coupling constant to be
\begin{equation}
  \label{coupling}
   g=\frac{k}{8\pi}\quad \textrm{with}\quad k=1,2,\ldots\, ,
\end{equation}
where we used that our anti-hermitian Lie-algebra generators, given below  (\ref{FUU}), are normalized
as  $\mathrm{tr}(T_MT_N) = -\fr{1}{2}\,\delta_{MN}$ 
and $\mathrm{tr}(\ww{T}_{\ww M}\ww{T}_{\ww N}) = -\fr{1}{2}\,\delta_{\ww{M}\ww{N}}$. Here we also assumed   
that there are no contributions from boundary terms in the gauge transformation of the CS-action 
\cite{Moore:1989yh}. We will comment on this in the following section.

The Lagrangian (\ref{labjm}) is formally written  in superspace. If the superfield equations imply 
the constraint 
equations this Lagrangian would provide an off-shell superspace formulation at 
least in the sense that 
on-shell the field content describes the ABJM model. Independently of this formal observation, 
the superfield expansions (\ref{eq:drec}) implies that the lowest component in the
$\theta$-expansion of this Lagrangian
gives directly the ABJM Lagrangian and that the lowest component of the superfield 
equations (\ref{csconf}), (\ref{confeom}) give 
the associated component equations or vice versa, respectively. We emphasized already 
that in our formalism the superfield 
expressions are often formally identical to their component field counterparts. This fact
will be convenient also in the following sections.

\section{$\ca{N}=8$ Enhancement, Monopole Operators}

\subsection{General Properties of Monopole Operators}

With the formulation of the  ABJM model and its proposed AdS/CFT 
duality relation to $M2$ branes sitting at the singularity of 
$\mathbb{C}^4/\mathbb{Z}_k$ it was also argued 
that for CS-coupling $k=1,2$, supersymmetry should be enhanced to $\ca{N}=8$ 
and that monopole operators might play a crucial role in this enhancement \cite{Aharony:2008ug}.
Before we implement this structure  into our formalism we discuss
some general properties of these monopole operators.

Monopole operators were first introduced by 't Hooft for $3D$- and $4D$- gauge theories
to define an alternative criterion for confinement \cite{tHooft:1977hy}.
Therefore monopole operators are often called 't Hooft operators\footnote{Though the concept of 't Hooft 
operators is more general, as they might also be non-topological \cite{Kapustin:2005py}.}. 
We rephrase here some of the 
illustrative arguments given in \cite{tHooft:1977hy}. The basic idea of 't Hooft was
to introduce an operator that creates/annihilates topological quantum numbers, i.e.\ magnetic 
fluxes. For solitons (monopoles, vortices)  it is known that the associated fields can be 
written as pure gauge configurations which are singular at the position of the soliton. 
Following this observation one can define an operator that acts as a gauge transformation 
$\Omega^{[x_0]}(\varphi)\in G$
which is singular at the insertion point $x_0$, $\varphi$ is the spatial angular coordinate with center 
$x_0$. Such gauge transformations have a nontrivial monodromy (winding), given by a center element
of the gauge group $G$, for example if $G=SU(N)$:
\begin{equation}
  \label{mop1}
  \Omega^{[x_0]}(2\pi)= e^{2\pi i\,n/N}\,\Omega^{[x_0]}(0)\ \ ,\ \  e^{2\pi i\,n/N}\in Z(G)\ ,
\end{equation}
with $n$ being an integer. Such a prescription defines an operator 
whose action can be described in a Hilbert space basis $|A_i(\vec x),H(\vec x)\rangle$,
which are eigenstates of the (spatial) gauge field operator $\hat A_i$ and matter field operators $\hat H$
(e.g.\ Higgs field, etc.) with given eigenvalues $A_i(x)$ and $H(x)$. The 
so defined monopole operator is then given by
\begin{equation}
  \label{mop2}
  \ca{M}( x_0) |A_i(\vec x),H(\vec x)\rangle =|A_i^{\Omega^{[x_0]}}(\vec x),H^{\Omega^{[x_0]}}(\vec x)\rangle\ ,
\end{equation}
where $A_i^{\Omega^{[x_0]}}(\vec x)$, $H^{\Omega^{[x_0]}}(\vec x)$ are the fields obtained by the 
gauge transformation described in (\ref{mop1}). From this representation it is clear that 
that the monodromy (\ref{mop1}) requires the matter fields denoted by $H$ to be invariant 
under the center $Z(G)$ of the gauge group,  for example to transform in the adjoint representation. 
Otherwise the states obtained by the action 
of the monopole operator (\ref{mop2}) are multi-valued and the monopole operator cannot be defined
directly in this way. For $Z(G)$-invariant matter it was shown in \cite{tHooft:1977hy} that this 
prescription defines 
a local operator and that the class of gauge transformations which have nontrivial monodromy 
(\ref{mop1}) in case that the insertion point is encircled and trivial monodromy otherwise is generated
by the Cartan subalgebra. Further it was shown in \cite{tHooft:1977hy} that operator insertion of
such kind can be described as the prescription of appropriate singularities in the elementary
fields when integrated in the path integral (also for $Z(G)$ variant matter).

It turns out to be a general method to define \emph{local} operators by requiring the elementary 
fields in the path integral to have certain singularities at the insertion point, a  point of view
 very much appreciated in \cite{Kapustin:2005py}. In a $CFT$, such as the ABJM model, one has in addition the operator-state 
correspondence, so that local operators can be described (in radial quantization) by states corresponding 
to the specified boundary conditions of the elementary fields. Monopole operators are understood 
as singularity prescriptions in the gauge field\footnote{For symmetry reasons also the matter fields might have 
corresponding singularity prescriptions.} which create magnetic $U(1)$ flux embedded in 
the gauge group under consideration. With this understanding one can compute perturbatively quantum numbers
(expectation values) 
for these operators by simply expanding the quantum fields around the specified singularities. This 
was done for different models in \cite{Borokhov:2002ib,Borokhov:2002cg,Borokhov:2003yu}, whereas considerable 
effort was necessary to circumvent 
the problem of strong coupling in the ABJM model \cite{Kim:2009wb,Benna:2009xd}. This procedure is in complete analogy 
to quantum computations for solitons, where the theory under consideration is quantized in a soliton background
and posses surprising effects, see 
\cite{Goldhaber:2004kn,Rebhan:2004vn,Rebhan:2006fg} for example.

The explicit prescription for monopole operator insertion in a three-dimensional gauge theory 
is as follows. The gauge field is supposed to have a Dirac monopole singularity at the 
insertion point, i.e
\begin{equation}
  \label{dirac}
  A_{N/S} = \frac{H}{2}\frac{\pm 1 -\cos{\theta}}{r}d\varphi\ ,
\end{equation}
such that it produces magnetic flux in a $U(1)$ subgroup through a sphere surrounding the insertion 
point in three dimensions\footnote{\label{ft}The signs correspond to N-/S-pole of the surrounding sphere. The 
surrounding sphere and the associated flux quantization, see below, may be best understood in the 
euclidean setting or in the radial 
quantization picture, where $\mathbb{R}^3\ \rightarrow\ S^2\times\mathbb{R}$.}. Hence such an 
operator prescription creates/annihilates 
topological quantum numbers.
It was shown in \cite{Goddard:1976qe} (for static monopoles in $4D$ gauge theories which is equivalent to the
situation 
considered here) that $H$ has to satisfy the quantization condition $e^{2\pi\,i\,H}=\unit_{G}$ and
therefore  is an 
element of the Cartan subalgebra $\mathfrak{t}\subset \mathfrak{g}$ of the form
$H=\mathrm{diag}(q_1,\ldots,q_N)$. The integers $q_i$ are the fluxes in the $U(1)$ subgroups. 
The GNO charges ${}^Lw=(q_1,\ldots,q_N)$, form a 
highest weight state 
of the GNO - or Langlands dual group ${}^LG$, where Weyl-reflection can be used to choose $q_1\geq \ldots\geq q_N$,
see also \cite{Kapustin:2005py,Kapustin:2006pk}. Therefore monopole operators are classified by irreps of
the dual gauge group ${}^LG$ and come in representations specified by the highest weight ${}^Lw$.

In the presence of a Chern-Simons term with CS-level $k$ the GNO charges are of the form $q_i=k\,n_i$, with 
integers $n_i$, as will be seen in a specific situation below. Taking also into account that 
${}^L(G_1\times G_2)= {}^LG_1\times {}^LG_2$ and that $U(N)$ is selfdual, i.e.\ ${}^LU(N)=U(N)$ one finds that
for the $U(N)\times {U}(\ww N)$ ABJM model monopole operators are in the representation $({}^Lw,{}^L\ww w)$
with $q_i=k\,n_i$ and $\ww q_i= - k\,\ww n_i$. In the following,  monopole operators of the form 
$\ca{M}^{ab}_{\ww a \ww b}$ are of particular interest, i.e.\ we choose\footnote{Due to the dynamics of the ABJM 
model the fluxes ${}^Lw$, ${}^L\ww w$ have to satisfy certain constraints, see e.g.\ \cite{Kim:2010ac} which 
is the case with 
the here given choice which also implies $\ww N=N$.} ${}^L\ww w = \overline{{}^L w}$, the conjugate 
representation. Then one has the possible weights
\begin{equation}
  \label{weights}
  {}^L w=(2,0,\dots,0)\quad \textrm{and} \quad {}^L w=(1,1,0,\dots,0)\ ,
\end{equation}
the associated representations have Young tableaux {\tiny\yng(2)} and {\tiny\yng(1,1)} 
\cite{Klebanov:2009kp,Klebanov:2009sg}. The first weight in (\ref{weights}) is possible for 
$k=1,2$ with $n_1=2,1$ and the associated monopole operator is in the 
$({\bf N}^2_{\mathrm{sym}},{\bf \bar N}^2_{\mathrm{sym}})$
representation, whereas the second weight allows only for $k=1$ with $n_1=n_2=1$ and the 
monopole operator is in the  $({\bf N}^2_{\mathrm{asym}},{\bf \bar N}^2_{\mathrm{asym}})$
representation.

The  prescription given above defines operator insertions which create topological 
quantum numbers, i.e.\ magnetic fluxes. Such a prescription is very different from the definition 
of a local (composite) operator as a polynomial local function of the elementary fields
and in general such a description will not be   available. Nevertheless it would 
be of great interest to have a more explicit formulation for such operators, especially
with regard to the conjectured dualities between different gauge theories. It is believed that such dualities
have the same origin as for example the well understood duality between the sine-Gordon and massive Thirring model
\cite{Coleman:1974bu,Mandelstam:1975hb} which arises from reformulating the model in 
terms of local operators which are non-polynomial
non-local functionals of the elementary fields and create topological quantum numbers. 
So far  such an explicit formulation for higher dimensional 
($D > 2$) (nonabelian) gauge theories has not been found. In three dimensions such a duality is mirror symmetry, which was 
first proposed in \cite{Intriligator:1996ex}\footnote{Another example is $S$-duality for $\ca{N}=4$
SYM and it was shown recently that  't Hooft and Wilson operators are related under under $S$-duality
 \cite{Kapustin:2005py,Kapustin:2006pk}.}. In the following we give a step in the direction of an explicit 
description of such dual degrees of freedom in terms of monopole operators. 
The reason why this is possible is susy enhancement which is supposed to be triggered by monopole operators
and thus results in very specific conditions which allow to specify these operators to a certain extent.

Before implementing monopole operators into the structure of superspace constraints we mention an 
example where monopole operators can  be constructed explicitly 
\cite{Moore:1989yh,Gawedzki:2001ye,Itzhaki:2002rc}, without going into 
details. For this we come back to idea of singular gauge transformations as described in the beginning 
of this section.
In the case of pure Chern-Simons theory (perhaps with matter invariant under the center 
$Z(G)$) one can define an operator insertion by defining a singular gauge transformation at the 
(spacial) insertion point. This carves out a tube of space time and as mentioned below (\ref{coupling}),
the CS-action has boundary contributions from a gauge transformation which gives a nonvanishing contribution
for singular gauge transformations even if the tube shrinks to zero. This change in the action
can be interpreted as an operator insertion. For $G=U(N)$ this operator is
a Wilson line of the form,
\begin{equation}
  \label{csop}
  \ca{M}(x)=\frac{1}{d(R)}\, \mathscr{P}\exp{(-\int^{x}_{\infty}A^{R})}\ .
\end{equation}
Here 
$d(R)$ is the dimension of the representation $R$ in which the connection $A$ is given 
and the highest weight of $R$ is  $w_R = k w_{\mathrm{fund}}$ with $w_{\mathrm{fund}}$ being
the highest weight of the fundamental representation ${\bf N}$ of $U(N)$. Under gauge transformations
$\ca{M}$ transforms as $R(g)(x)\ca{M}R(g^{-1})(\infty)$ and thus w.r.t.\ genuine 
gauge transformations ($g(\infty)=\unit$)  the monopole operator transforms in the $({\bf N}^k_{\mathrm{sym}})$
representation. Due to the special dynamics of pure CS-theory this operator is local, i.e.\ it depends only on the 
endpoint of the path, 
but this 
construction does not work in the case charged matter is present. However, the principal structure
is an appealing guideline.

\subsection*{Monopole Operator Superfields}

In the following we investigate the possible supersymmetry enhancement of the $\ca{N}=6$ ABJM model 
to $\ca{N}=8$ supersymmetry for the gauge groups $U(N)\times U({\widetilde N})$ with a priori arbitrary 
$N, \widetilde{N}\geq 2$. We start from our basic description of the model through the 
superspace constraints (\ref{dc2}), (\ref{defW}) with the deformation potential (\ref{Wconf}). 
The principle idea is to realize the additional $\ca{N}=2$ susy as infinitesimal internal fermionic symmetry 
of the superspace constraints. The superspace description is on-shell, and thus any 
 additional 
supersymmetry obtained in this way will be, a priori, an (infinitesimal) symmetry of the 
classical e.o.m.\ only. This is a situation
well known from two dimensional models with hidden symmetries, or even in four dimensions \cite{Wolf:2004hp}.
 Nevertheless, there
should exist currents associated with the hidden symmetry which are conserved on-shell, i.e.\ 
dynamically. We assume the existence of a (composite) monopole operator superfield to be able to
formulate these additional symmetries 
and derive superspace constraints which this operator has to satisfy in order to obtain susy enhancement.

In the following it will be convenient to use the more compact notation as employed before,
with a single index labelling the matter field representation $R$. 
Only when necessary translate the expressions to the explicit notation as described in 
(\ref{schnabel}).
The starting point is the transformation behaviour for the bosonic
superfield under the additional supersymmetry, which we  define as follows:
\begin{equation}
  \label{dbos}
  \delta\Phi^{i\, a}:=\epsilon^{\alpha} \ca{M}^{ab}\,\bar\Psi_{\alpha\,b}^i \ \ ,\ \ 
    \delta\bar\Phi_{\,i\, a}:=-\bar\epsilon^{\,\alpha}{\bca{M}}_{ab}\,\Psi_{\alpha\,i}^{b}
\end{equation}
where $\epsilon^{\alpha}$ is a complex anti-commuting constant spinor and $\ca{M}^{ab}=:({\bca{M}}_{ab})^\ast$ 
is the proposed monopole operator superfield.
It is supposed to be a \emph{local} composite superfield but in general non-polynomial in and a
nonlocal functional of the elementary 
superfields. 
It is necessary that the superfields $\Phi^i$ transform 
into superfields $\bar\Psi_{\alpha}^i$ to obtain a transformation different from the 
original $\ca{N}=6$ susy. Therefore one has to assume the existence 
of a monopole operator so that the transformation (\ref{dbos}) is gauge covariant. 

Taking the canonical dimension for the 
new susy parameter to be the standard one, i.e.\ $[\epsilon^{\alpha}]=[\theta^{\alpha\,ij}]=-\frac{1}{2}$,
one sees that the monopole operator has canonical dimension zero, i.e.\ $[\ca{M}^{ab}]=0$. 
Given that the monopole operator should have 
canonical dimension zero and that the extra susy should commute with the $SU(4)$ R-symmetry 
(the susy parameter $\epsilon$ is an  $SU(4)$ singlet), the transformation 
rule (\ref{dbos}) is the only conceivable one. As a 
matter of fact, this is the only \emph{ansatz} that we make, all other relations will be derived. This 
clearly  shows the effectiveness of the   formalism developed here.

As mentioned, the monopole operator has to compensate the different gauge
transformation properties of the elementary fields in (\ref{dbos}), which for the non-abelian part of
the gauge group determines the representation (indices) it carries. In addition, for possible gauged $U(1)$
factors the monopole operator has to carry appropriate $U(1)$ charges so that (\ref{dbos}) is gauge covariant.
We will discuss this point in more detail below.

\subsection{$U(2)\times U(2)$: Fake Monopole Operators}

Having defined the transformations for the bosonic superfields we have to determine the transformations
of the residual elementary superfields such that the superfield constraints (\ref{defW}), (\ref{dc2}) 
are invariant under these transformations.
This will necessarily impose also conditions on the monopole operator superfield. In this subsection 
we will assume that the monopole operator is covariantly constant, 
an assumption which has been considered also in \cite{Kwon:2009ar} and \cite{Gustavsson:2009pm}. 
We will show that this 
eventually restricts the gauge group to be $U(2)\times U(2)$ (or $SU(2)\times SU(2)$). Later we will
relax this condition and derive superspace constraints for the 
monopole operator without any obvious restriction on the dimension/rank of the gauge group.  
The main point is that 
we do not allow any additional condition on the elementary component or superfields of the theory, i.e.\ 
for $\Phi^i,\Psi_{\alpha\,i}, \ca{A}_{\alpha\,ij},\ca{A}_{\alpha\beta}$, since this would change the theory. 
This is a major difference to the considerations in \cite{Gustavsson:2009pm} 
where numerous nontrivial conditions (called ``identities'') are imposed 
on the elementary fields. While finding nontrivial
solutions for these conditions represents a formidable task the very presence of
such conditions inevitably changes the original model.
\\

\noindent  
{\bf Matter Constraint.}
We first consider the matter constraint (\ref{dc2}) which for the 
transformation of the fields reads 
\begin{equation}
  \label{dmatter}
  \left\{\nabla_{\alpha\,ij}\delta\Phi^k+\delta\ca{A}_{\alpha\,ij}\cdot\Phi^k\right\}
             \big|_{{\bf \overline{20}}} \overset{!}{=}0\ .
\end{equation}
Thus one has to define $\delta\ca{A}_{\alpha\,ij}$ such that for the given transformation 
$\delta\Phi^k$ (\ref{dbos}) the ${\bf\overline{20}}$ contribution in
the product ${\bf \bar{6}}\otimes {\bf 4}={\bf\bar{4}}\oplus{\bf\overline{20}}$ in 
(\ref{dmatter}) vanishes. The remaining part, transforming in the ${\bf \bar{4}}$ defines
the enhanced susy transformation of the fermionic superfield $\Psi_{\alpha\,i}$, see (\ref{dc2}). 
Explicitly one finds with the complex conjugate of (\ref{DPsi}),
\begin{equation}
  \label{dmatter2}
  \delta\ca{A}_{\alpha\,ij}{}^a{}_d \,\Phi^{k\,d}
        -\epsilon^{\beta}\left(\nabla_{\alpha\,ij}\ca{M}^{ab}\,\bar{\Psi}_{\beta\,b}^{k} -
         \epsilon_{\alpha\beta}\,\ca{M}^{ab}\F{c}{e}{b}{d}\,\bar{\Phi}_{i\,c}\bar{\Phi}_{j\,e}\Phi^{k\,d}\right) \ \
         \overset{!}{\sim}\ {\bf \bar{4}}\ ,
\end{equation}
where we have used that for the conformal deformation potential (\ref{Wconf}) 
($W^k{}_{[i}\cdot\bar\Phi_{j]})_{b}=\F{c}{e}{b}{d}\,\bar\Phi_{i\,c}\bar\Phi_{j\,e}\Phi^{k\,d}$,
 up to terms transforming in the ${\bf\bar{4}}$.

At this point we impose the condition that the monopole operator is covariantly constant,
\begin{equation}
  \label{Mconst}
  \nabla_{\alpha\, ij}\ca{M}^{ab}=0\ .
\end{equation}
Covariant constancy is defined here with respect to the fermionic covariant derivative. We will discuss
in a moment the implications of the integrability conditions of (\ref{Mconst}) and show that as a consequence  
the monopole
operator is also covariantly constant w.r.t.\ the bosonic covariant (super) derivative. Imposing (\ref{Mconst})
it seems to be trivial to choose $\delta\ca{A}_{\alpha\,ij}$ such that the invariance condition 
for the matter constraint (\ref{dmatter2}) is satisfied, but there is an additional obstacle. 
The transformation of the gauge superfield $\ca{A}_{\alpha\,ij}$ has to conserve the reality condition 
(\ref{eq:1.3}), so that
\begin{equation}
  \label{dAf}
  \delta\ca{A}_{\alpha\,ij}{}^a{}_d=\epsilon_{\alpha}\,(\ca{M}\cdot f)^{a,ce}{}_d\bar\Phi_{i\,c}\bar\Phi_{j\,e}
                 +\fr{1}{2}\,\bar\epsilon_{\alpha}\,\varepsilon_{ijkl}
                (\ca{\bar M}\cdot f)_{d,ce}{}^a\Phi^{k\,c}\Phi^{l\,e}\ ,
\end{equation}
where we introduced the abbreviation $(\ca{M}\cdot f)^{a,ce}{}_d:=\ca{M}^{ab}\F{c}{e}{b}{d}$ and analogously
for the complex conjugate expression. The first term in (\ref{dAf}) is determined by the invariance condition 
(\ref{dmatter2}) whereas the second term is necessary to obey the reality condition (\ref{eq:1.3})
\footnote{There is still the freedom to add a  term  $\delta\ca{\ww A}_{\alpha\,ij}$ for which 
one has to assume that it satisfies 
$\delta\ca{\ww A}_{\alpha\,ij}\cdot\Phi^k\big{|}_{\overline{\bf 20}}=0$, but such a term does not 
play a role in the  
following considerations. We will comment on this in the succeeding section when we consider covariant 
non-constant monopole operators.}.
Inserting the
transformation (\ref{dAf}) back into the invariance condition (\ref{dmatter2}) results in the following condition:
\begin{equation}
  \label{Mfcon}
  (\ca{\bar M}\cdot f)_{d,ce}{}^a\Phi^{k\,c}\Phi^{l\,e}\Phi^{k\,d}\  \overset{!}{\sim}\ {\bf \bar{4}}
  \qquad \Leftrightarrow\qquad    (\ca{\bar M}\cdot f)_{d,ce}{}^a\  \overset{!}{=}\  
(\ca{\bar M}\cdot f)_{[dce]}{}^a \ .
\end{equation}
 With the explicit solution (\ref{schnabel}) for  $\F{a}{b}{c}{e}$ and employing the explicit 
notation of  (\ref{schnabel}) for the monopole operator one finds a unique solution 
to this condition:
\begin{equation}
 \label{fakeM}
  N=\widetilde{N}=2\ \ \ \mathrm{and}\ \ \  \ca{\bar M}^{\ww a \ww b}_{a b}
        \ \sim \ \varepsilon^{\ww a \ww b}\varepsilon_{ab}\ .
\end{equation} 
The only other possibility is the trivial case $ N=\widetilde{N}=1$. Before we discuss the possible proportionality 
factor for the monopole operator in (\ref{fakeM}) we investigate the invariance of gauge field constraint (\ref{defW}).
Here we just notice, that the monopole operator (\ref{fakeM}) transforms in the 
$({\bf N}^2_{\mathrm{asym}},{\bf N}^2_{\mathrm{asym}})$.
\\

\noindent
{\bf Gauge Field Constraint.}
We have now defined the enhanced susy transformation for the scalar and fermionic vector superfields
$\Phi^i, \ca{A}_{\alpha\,ij}$ (and for the fermionic superfield $\Psi_{\alpha\,i}$, though we did not write it explicitly),
such that the matter constraint is invariant with a covariantly constant monopole operator. This imposed the 
restrictions (\ref{fakeM}) on  gauge group and the monopole operator. The invariance of the gauge field constraint
(\ref{defW}) requires
\begin{equation}
  \label{dgauge}
  \nabla_{\alpha\,ij}\,\delta\ca{A}_{\beta\,kl}+\nabla_{\beta\,kl}\,\delta\ca{A}_{\alpha\,ij}\ \overset{!}{=}\ 
   i\,(\varepsilon_{ijkl}\,\delta\ca{A}_{\alpha\beta}+\varepsilon_{\alpha\beta}\varepsilon_{mij[k}\,\delta W^m{}_{l]})\ ,
\end{equation}
where $\delta W^m{}_{l}$ is the susy transformation of the conformal deformation potential 
(\ref{Wconf}) obtained from the transformation of the scalar superfields (\ref{dbos}). Inserting the 
transformation (\ref{dAf}) and using the matter constraint (\ref{dc2}) one obtains 
for the transformation of the bosonic super vector field
\begin{equation}
  \label{dAb}
  \delta\ca{A}_{\alpha\beta}{}^a{}_d=
           \fr{1}{2}\,  (\ca{M}\cdot f)^{a,ce}{}_d\,\bar\Phi_{k\,c}\,\epsilon_{(\alpha}\bar\Psi^k_{\beta)e}
             +  \fr{1}{2}\,  (\ca{\bar M}\cdot f)_{d,ce}{}^a\,\Phi^{k\,c}\,\bar\epsilon_{(\alpha}\Psi^e_{\beta)k}\ ,
\end{equation}
where we have used the conditions (\ref{Mconst}), (\ref{fakeM}) for the 
monopole  operator and did not encounter any further restrictions. \\

\noindent
\underline{\bf ${ SU(2)\times SU(2)}$:}  The assumption that the monopole operator is covariant 
constant resulted in the restriction\footnote{The mentioned trivial solution $N=\widetilde{N}=1$ refers to 
the $U(1)\times U(1)$ case which was considered in detail in \cite{Aharony:2008ug} and \cite{Kwon:2009ar}. 
The considerations  become become rather trivial in this case since $\F{a}{c}{b}{d}=0$ for $U(1)\times U(1)$.} 
$N=\widetilde{N}=2$ and the particular form for the monopole operator 
as given in (\ref{fakeM}). For the gauge group $SU(2)\times SU(2)$ there is no local $U(1)$ transformation
which has to be compensated by the operator $\ca{M}^{ab}$ in the transformation rules (\ref{dbos}), (\ref{dAf}) and 
(\ref{dAb}) so that the proportionality factor in (\ref{fakeM}) is an arbitrary number which can be absorbed in the 
susy parameter $\epsilon$. Thus one has
\begin{equation}
  \label{su2M}
  \ca{M}^{ab}_{\ww a\ww b}=\varepsilon^{ab}\varepsilon_{\ww a\ww b}\ .
\end{equation}
Clearly this ``operator'' is  covariantly constant, i.e.\ satisfies (\ref{Mconst}), and the gauge field constraint 
(\ref{defW}) is consistent with the fact that this $SU(2)\times SU(2)$ invariant is also covariantly constant w.r.t.\ the bosonic 
covariant derivative. 
This is not a monopole operator in the sense discussed above (it does not create any
 magnetic flux), 
but rather shows that for the 
gauge group $SU(2)\times SU(2)$ the supersymmetry is ``kinematically'' enhanced to $\ca{N}=8$.  This case 
actually describes the BLG model \cite{Bagger:2007jr, Gustavsson:2007vu, Aharony:2008ug}. 
In \cite{VanRaamsdonk:2008ft} 
a $SU(2)\times SU(2)$ formulation of the BLG model with manifest $SO(8)$ R-symmetry was given but at the
cost of an additional condition on the matter fields, which reads for the scalar fields as 
$\varepsilon_{\ww a\ww b}\, \bar{X}^{I\,\ww b}{}_a\,\varepsilon^{ab}= X^{I\,b}{}_{\ww a}$, $I=1,\ldots 8$,  
and similar for the fermions.
In the equivalent formulation of ABJM, as given here, no such constraint is present, but therefore the manifest 
R-symmetry is only $SU(4)$. The fields in the two different descriptions are related as follows,
\begin{equation}
  \label{BLABJM}
  \Phi^{i\,a}{}_{\ww b}=X^{i\,a}{}_{\ww b}+i\, X^{i+4\,a}{}_{\ww b}\ \ ,\ \ 
   \bar\Phi_i{}^{\ww b}{}_{a}=\bar X_i{}^{\ww b}{}_{a} - i\, \bar{X}_{i+4}{}^{\ww b}{}_{a}\ ,\ \ i=1,\ldots, 4 \ ,
\end{equation}
which shows that the ABJM formulation resolves the mentioned condition 
at the price of loosing manifest $SO(8)$ R-symmetry according to
\begin{equation}
  \label{so8su4}
  SO(8)\ \rightarrow\ U(1)\times SU(4)
\end{equation}
with remaining manifest $SU(4)$ $R$-symmetry. Furthermore, in  the case of an $U(2)\times U(2)$ gauge group the 
$U(1)$ factor in (\ref{so8su4}) is also gauged. This is the case that we discuss next.\\

\noindent
\underline{\bf $U(2)\times U(2)$:} 
The $U(1)$ sectors of the the Lagrangian (\ref{labjm}) are of a very particular form. The bifundamental
action of the covariant derivative (\ref{dbifu}) implies that the matter fields couple exclusively
to the difference of the two $U(1)$ gauge fields\footnote{In accordance with our normalization of 
the generators, see below (\ref{coupling}), the $U(1)$ generators are given by $T^0=\frac{i}{\sqrt{2N}}\unit$ 
with $N=2$ in the considered case.},
\begin{equation}
  \label{Ab}
  \ca{A}_{\alpha\beta}^{\mathrm{bar}} = \fr{1}{2}(\mathrm{tr}\ca{A}_{\alpha\beta}-\mathrm{tr}\ca{\ww A}_{\alpha\beta})\ ,
\end{equation}
where the superscript ``$\mathrm{bar}$'' stands for baryonic. Thus the 
 $U(1)$ factor described in (\ref{so8su4}) is the baryonic $U(1)_b$ gauge symmetry. The matter fields
$(\Phi^i,\Psi_i)$ have $U(1)_b$ charge $+1$ while the c.c. thereof have $U(1)_b$ charge $-1$. Denoting the 
opposite combination of the $U(1)$ gauge fields, i.e.\ the diagonal $U(1)$ in $U(2)\times U(2)$,   by
 $\ca{A}_{\alpha\beta}^{\mathrm{diag}}=\fr{1}{2}(\mathrm{tr}\ca{A}_{\alpha\beta}+\mathrm{tr}\ca{\ww A}_{\alpha\beta})$,
the $U(1)$ sector of the CS-Lagrangian in (\ref{labjm}) writes as
\begin{equation}
  \label{lu1}
  d\mathrm{Vol}\,\ca{L}^{U(1)}_{CS} = - \frac{k}{4\pi}(\ca{A}^{\mathrm{bar}}\wedge d\ca{A}^{\mathrm{diag}} 
 + \ca{A}^{\mathrm{diag}}\wedge d\ca{A}^{\mathrm{bar}})\ .
\end{equation}
The gauge field $\ca{A}^{\mathrm{diag}}$ appears only at this place and its variation 
imposes a flatness condition for the  baryonic $U(1)$ connection. Explicitly, the 
e.o.m.\ (\ref{csconf}) for the $U(1)$ sectors write as
\begin{equation}
  \label{u1eom}
  \ast \ca{F}^{\mathrm{bar}} = 0\  \ , \ \  \frac{k}{2\pi}\ast \ca{F}^{\mathrm{diag}} = -j^{\mathrm{bar}}\ ,
\end{equation}
where $j^{\mathrm{bar}}$ is the current associated with the $U(1)_b$ symmetry. It has been shown 
in several places, e.g.\ \cite{Aharony:2008ug}, that the diagonal $U(1)$ field strength 
$\ca{F}^{\mathrm{diag}}=d\ca{A}^{\mathrm{diag}}$ which appears as a Lagrange multiplier (up to a surface term)
in (\ref{lu1}) can be considered as a fundamental field and treated in for a dual $2\pi$-periodic 
scalar $\tau$,
such that the flatness condition for the baryonic connection is expressed as 
$\ca{A}^{\mathrm{bar}}=\frac{1}{k} d \tau$.

From the point of view of the classical dynamics the $U(1)$ factors are irrelevant, only the 
baryonic connection $\ca{A}^{\mathrm{bar}}$ couples to the matter but due to its flatness  
 it can be gauged away locally. The ``dynamics'' of the diagonal field strength (\ref{u1eom}), 
appearing as a Lagrange multiplier, is just the attachment of magnetic ``flux'' in the diagonal $U(1)$ 
factor to the ``electric'' $U(1)_b$ current, an effect well known in the presence of a CS-field 
\cite{Dunne:1998qy}. This e.o.m.\ implies that magnetic flux and
$U(1)_b$ charge obey a Dirac-quantization condition according to (see also footnote \ref{ft})
\begin{equation}
  \label{chargequant}
  \frac{k}{2\pi} \int_{S^2}  \ca{F}^{\mathrm{diag}} = k\,n =  \int_{S^2}\ast j^{\mathrm{bar}} = q_{\mathrm{bar}} \ ,
\end{equation}
which is understood to be read for the lowest component of the superfield expressions. Though dynamically
irrelevant the  gauged $U(1)$ factors have an important impact on the moduli space of the 
theory, in particular on the possible vacua in the quantum theory. The additional identification
due to the $U(1)$ gauge symmetry reduces the naive moduli space 
$\mathbb{C}^4$ to the orbifold $\mathbb{C}^4/\mathbb{Z}_k$ \cite{Aharony:2008ug}.

We now describe  the effect for the monopole operator of the just discussed $U(1)$ factors. In the case that the 
gauge group is $U(2)\times U(2)$ the operator (\ref{su2M}) cannot be the whole answer,
it is neither covariantly constant nor does it carry the right $U(1)_b$ charge. One therefore sets 
$\ca{M}^{ab}_{\ww a\ww b} = T \varepsilon^{ab}\varepsilon_{\ww a\ww b}$, so that covariant constance implies 
for the allowed (\ref{fakeM}) pre-factor
\begin{equation}
  \label{u2u2M}
   \nabla_{\alpha\,ij}  \ca{M}^{ab}_{\ww a\ww b} = 0\quad \Rightarrow\quad 
  D_{\alpha\,ij}T + 2\, \ca{A}^{\mathrm{bar}}_{\alpha\,ij}\,T=0\ .
\end{equation}
The integrability condition for this equation gives with the gauge constraint (\ref{defW})
\begin{equation}
  \label{icm}
   \nabla_{\alpha\beta}  \ca{M}^{ab}_{\ww a\ww b} = 0\quad \Rightarrow\quad 
  \del_{\alpha\beta}T + 2\, \ca{A}^{\mathrm{bar}}_{\alpha\beta}\,T=0\ ,
\end{equation}
where for the particular form of $\ca{M}^{ab}_{\ww a\ww b}$ the conformal
deformation potential (\ref{Wconf}) does not give any contribution. Thus also in the $U(2)\times U(2)$
case covariant constance w.r.t to the fermionic connection implies covariant constance in the 
usual sense. This in fact results in yet another integrability condition, i.e.\ the existence of 
non-trivial solutions to (\ref{icm})
requires $\ca{F}^{\mathrm{bar}} = 0$, which is satisfied due to the e.o.m.\ (\ref{u1eom}).

Solutions to equations of the form (\ref{u2u2M}), (\ref{icm}) are given by local,
i.e.\ path independent, Wilson lines which are abelian  here. To 
implement this in superspace an unimportant but necessary little formality has to be introduced. 
The super connections $\ca{A}_A = (\ca{A}_{\alpha\beta},\ca{A}_{\alpha\,ij})$, as usual in supersymmetric
theories, is defined w.r.t to the non-holonomic basis $D_A = (\del_{\alpha\beta},D_{\alpha\,ij})$. The 
relation to the holonomic coordinate basis $\del_M = (\del_{\alpha\beta},\del_{\alpha\,ij})$ is
given by a super vielbein, i.e.\ 
  $\del_{M} = e_M{}^A D_A$.
For more details on superspace Wilson lines see \cite{Karp:2000mn}. For the purpose of the present consideration,
there is no need to go into further detail\footnote{For completeness, the super vielbein  is given by
$e_M{}^A = \begin{bmatrix}
               \delta^{\gamma}_{(\alpha}\delta^{\delta}_{\beta)} & 0 \\
                -i\theta^{(\delta}_{ij}\delta^{\gamma)}_{\alpha} & \delta^{\gamma}_{\alpha}\delta^{kl}_{ij}
            \end{bmatrix}
$.}. The solution to (\ref{u2u2M}), (\ref{icm}) is then given by
\begin{equation}
  \label{TWl}
  T(x,\theta) = 
    \exp\,\{-2\, \int_{\tau_{\infty}}^{\tau_{x}}d\tau\, \dot{z}^M(\tau)\, e_M{}^A\ca{A}^{\mathrm{bar}}_A(z(\tau))\}\ ,
\end{equation}
where $z^M(\tau)=(x^{\alpha\beta}(\tau),\theta^{\alpha\,ij}(\tau))$ is a path in superspace with 
$z^M(\tau_x)=(x^{\alpha\beta},\theta^{\alpha\,ij})$ being the insertion point and otherwise to be specified in 
a moment. The lowest
component of the superfield $T$ is just an ordinary Wilson line, i.e.\ $T|_{\theta = 0}=e^{-2\int_{\ca{C}}A}$.
In this form the superfield $T$ satisfies the equation
\begin{equation}
  \label{Tequ}
  \dot{z}^M\, e_M{}^A(D_A\ T + 2\,\ca{A}^{\mathrm{bar}}_{A}\, T)_{\tau=\tau_x} = 0\ ,
\end{equation}
and under a $U(1)_b$ super-gauge transformation $\delta\ca{A}^{\mathrm{bar}}_{A} = - \frac{1}{2}D_{A}(\mathrm{tr}\,\Omega-\mathrm{tr}\,\ww\Omega)=:-D_{A}\Omega^{\mathrm{bar}}$ behaves as
\begin{equation}
  \label{Tgauge}
  T\ \ \rightarrow\ \ e^{\,2\, (\Omega^{\mathrm{bar}}(\tau_x)-\Omega^{\mathrm{bar}}(\tau_{\infty}))}\ T\ .
\end{equation}
The requirement that equation (\ref{Tequ}) implies  (\ref{u2u2M}), (\ref{icm}) is equivalent to 
the requirement that the operator $T$ is local, i.e.\ $T=1$ for any closed path and thus $\dot{z}^M$
can be arbitrary. This is guaranteed by the e.o.m.\ (\ref{u1eom}) which implies the already below (\ref{icm})
mentioned integrability condition. The requirement that the monopole operator has $U(1)_b$ charge $2$,
so that the transformation (\ref{dbos}) is gauge covariant, requires that 
$x^{\alpha\beta}(\tau_{\infty})\rightarrow \infty$ so that the monopole operator has charge $+2$ under genuine
gauge transformations for which then $\Omega^{\mathrm{bar}}(\tau_{\infty}) = 0$, a structure similar to 
(\ref{csop}).

We thus have for the $U(2)\times U(2)$ model the monopole operator superfield
\begin{equation}
  \label{u2u2mop}
  \ca{M}^{ab}_{\ww a\ww b} = T(x,\theta)\, \varepsilon^{ab}\varepsilon_{\ww a\ww b} \ ,
\end{equation}
with $T$ given in (\ref{TWl}). At zero order in $\theta$ this agrees with the result of \cite{Kwon:2009ar}.
The obtained monopole operator is in the 
$({\bf{N}}^2_{\mathrm{asym}},{\bf N}^2_{\mathrm{asym}})$ representation, a puzzling fact as we discuss in a moment.
In \cite{Kwon:2009ar} it is also argued that this monopole operator exists for $k=1,2$, but our 
general discussion below (\ref{weights}) indicates that it exists actually only for $k=1$. The 
criterion given in \cite{Kwon:2009ar} is perhaps not specific enough.

One may ask if covariant constancy w.r.t.\ the super-derivative (\ref{Mconst}) is a stronger (or too strong)
requirement compared to covariant constancy w.r.t.\ the bosonic connection, as it was implemented 
in an ordinary space-time approach in \cite{Kwon:2009ar}. To answer this question we look at 
the proposed enhanced $R$-symmetry
current \cite{Benna:2009xd,Klebanov:2009sg}. Employing the compact notation the enhancement current 
of \cite{Benna:2009xd,Klebanov:2009sg} can be 
promoted to a superfield expression in the form
\begin{equation}
  \label{kleb}
  \ca{J}^{ij}_{\alpha\beta}=i\,\ca{\bar M}_{ab}\{2\, \Phi^{a[i}\, \nabla_{\alpha\beta}\, \Phi^{j]b}
   -\fr{1}{4}\varepsilon^{ijkl}\, \Psi^a_{k(\alpha}\Psi^b_{\beta)l}\}\ .
\end{equation}
This current can be understood as the enhanced current superfield, with the enhanced $R$- and susy-current
at the lowest and first order in the $\theta$ expansion. For this current actually representing  new 
symmetries it 
has to be conserved (on-shell). Under the assumption  that
the monopole operator is covariantly constant w.r.t.\ to the bosonic 
(or ordinary space-time) connection this requirement leads to the same condition as given in (\ref{Mfcon}) for two 
independent reasons. First, using the e.o.m.\ (\ref{confeom}) one finds that the current is conserved only 
if  (\ref{Mfcon}) is satisfied. Second, $\nabla_{\alpha\beta}\ca{M}^{ab}=0$ implies an integrability condition
for the bosonic field strength, which together  with the e.o.m.\ (\ref{csconf}) again leads to the condition (\ref{Mfcon}).
Therefore it is clear that the assumption of a covariantly constant monopole operator is consistent only 
for $U(2)\times U(2)$ gauge group. 

However, as mentioned before the  monopole operator (\ref{u2u2mop}) gives rise to a puzzle. 
The current (\ref{kleb}) is a dimension 2 conformal primary and thus as a vector has to be conserved
in a unitary $CFT$. For this to be true the monopole operator has to have dimension zero, also in the 
quantized theory. This was proved in \cite{Benna:2009xd} for the case of a monopole operator in the 
$({\bf{N}}^2_{\mathrm{sym}},{\bf N}^2_{\mathrm{sym}})$, i.e.\ corresponding to the first weight in (\ref{weights}).
Further it was argued in \cite{Klebanov:2009sg} that with such a monopole operator one can form the 
$20$ missing dimension 1 operators to match the SUGRA spectrum on $S^7$ for $k=1$. Given that 
there is an monopole operator in the $({\bf{N}}^2_{\mathrm{asym}},{\bf N}^2_{\mathrm{asym}})$ representation
(\ref{u2u2mop}), this would mean that one would double the number of enhancement currents (\ref{kleb}).
Further, one could build additional dimension 1 operators of the form $\bar\Phi_i\ca{M}\bar\Phi_j$ 
and their complex conjugates, which would give additional (too many) $10 + 10$ states. We therefore
assume that the dimension of the monopole operator (\ref{u2u2mop}) is not protected and therefore 
is not connected to susy enhancement in the full theory. We therefore call this operator
in the context of susy enhancement ``fake monopole operator''\footnote{We want to mention however, that 
monopole operators in the $({\bf{N}}^2_{\mathrm{asym}},{\bf N}^2_{\mathrm{asym}})$ representation were
considered in the context of mass deformations of the ABJM model \cite{Klebanov:2009kp}}.

\subsection{$U(N)\times U(N)$: Monopole Operator Superspace Constraints}

We have shown in the previous section, that the assumption that the monopole operator 
is covariantly constant is too restrictive. 
We now relax this assumption and derive superspace constraints for
the (composite) monopole operator superfield, analogously to the formulation of the 
constraint equations for the matter and vector superfields (\ref{dc2}), (\ref{defW}),
such that the supersymmetry is enhanced
to $\ca{N}=8$ without any restriction on the rank of the gauge group factors.
\\

\noindent
{\bf Matter Constraint.} We again start from our basic and only assumption, the 
transformation rule for the bosonic superfield (\ref{dbos}) which leads to the invariance condition 
(\ref{dmatter}), (\ref{dmatter2}) for the matter constraint. Invariance of the matter constraint, equ.
(\ref{dmatter}), implies that 
the $\bar\epsilon$-part of enhanced susy transformation for the fermionic 
connection $\delta{\ca A}_{\alpha\, ij}$ has to satisfy an algebraic condition whereas the $\epsilon$-part
can now be absorbed in the super-derivative of the monopole operator. Explicitly we have the 
algebraic condition
\begin{equation}
  \label{Aalg}
  \delta{\ca A}_{\alpha\,ij}{}^a{}_d = \delta_{\epsilon}{\ca A}_{\alpha\,ij}{}^a{}_d 
     + \fr{1}{2}\varepsilon_{ijkl}[\delta_{\epsilon}{\ca A}_{\alpha\,ij}{}^d{}_a]^\ast \ \ \mathrm{with}\ \
      \fr{1}{2}\varepsilon_{ijkl}[\delta_{\epsilon}{\ca A}_{\alpha\,ij}{}^d{}_a]^\ast\ \Phi^{k\,d}
     \ \overset{!}{\sim}\ {\bf \bar 4}\ ,
\end{equation}
where the first equation implements the reality condition (\ref{eq:1.3}). 

We now solve the algebraic condition (\ref{Aalg}) and then determine the super-derivative 
of the monopole operator superfield. To this end we introduce a  structure, 
which will be extremely useful following, 
in terms of the determinant of \emph{mesonic} operators:
\begin{equation}
  \label{mesop}
  X^i{}_j:=\mathrm{tr}\,\Phi^i\bar\Phi_j = \Phi^{i\,a}\bar\Phi_{j\,a}\ \ ,\ \ |X|:=\mathrm{det}(X^i{}_j)\ .
\end{equation}
The mesonic operators $X^i{}_j$ form a hermitian matrix and transform in the $({\bf 4},{\bf\bar 4})$ under 
the R-symmetry $SU(4)$. We collect a number of curious relations which are needed in the following in the 
appendix \ref{Apmes}. For the understanding of the main text we introduce here a part of this structure, where  
we use the abbreviations $\del_{i\,a}:=\frac{\del}{\del\Phi^{i\,a}}$ and 
 $\bar\del^{i\,a}:=\frac{\del}{\del\bar\Phi_{i\,a}}$ in the following. 
First, with the help of the determinant of the mesonic operators one can translate an $SU(4)$ index
of \emph{any} quantity into the $SU(4)$ index of a scalar superfield, i.e.\  
\begin{equation}
  \label{mes1}
  \ca{O}^i|X|=\Phi^{i\,a}\, {\cal O}^j\, \del_{j\,a}|X| \quad\quad \mathrm{and}\quad\quad 
   \ca{O}_i|X|=\bar\Phi_{i\,a}\,{\cal O}_j\,\bar\del^{j\,a}|X|\ .
\end{equation}
Second, we introduce the hermitian \emph{projection operators}
\begin{equation}
  \label{projectors}
  \ca{F}^a{}_b:=\fr{1}{|X|}\,\Phi^{i\,a}\,\del_{i\,b}|X|=\fr{1}{|X|}\,\bar\Phi_{i\,b}\,\bar\del^{i\,a}|X|
   \quad \mathrm{and} \quad \ca{P}^a{}_b:= \delta^a{}_b - \ca{F}^a{}_b \ ,
\end{equation}
which have the properties
\begin{equation}
  \label{projectors2}
  \ca{F}\cdot\ca{F} = \ca{F}\ ,\ \  \ca{P}\cdot \ca{P} =  \ca{P}\ ,\ \ 
    \ca{P}\cdot\ca{F}=0 \ \ \mathrm{and} \ \  \ca{F}\cdot\Phi^i=\Phi^i\ ,\ \ \ca{P}\cdot\Phi^i=0\ ,
\end{equation}
where the gauge indices are contracted in an obvious way, indicated by a dot. The last two
identities have a hermitian conjugated counterpart for the fields $\bar\Phi_i$. In these 
definitions appears the inverse of the mesonic determinant $|X|$. For generic 
configurations this is a well defined superfield, only for isolated points in the field 
configuration space it might be singular. In general we allow such operators in our 
considerations. The field configurations where these operators become singular might be of 
special interest but we leave this question for future studies. 

With the introduced structure we can write down the general solution for the 
algebraic constraint (\ref{Aalg}). According to (\ref{Aalg}) there exists a (composite) superfield 
$ G^{\,a}_{\alpha\beta\,m}$ in the ${\bf \bar 4}$ of $SU(4)$ such that
\begin{equation}
  \label{Asol}
  [\delta_{\epsilon}{\ca A}_{\alpha\,ij}{}^d{}_a]^\ast\,\Phi^{k\,d} =
    \bar{\epsilon}^\beta\,\varepsilon^{ijkm}\ |X|\ G^{\ \ a}_{\alpha\beta\, m} \ ,
\end{equation}
where for convenience we pulled out a factor $|X|$, which is pure convention (c.f. the 
remarks above regarding the existence of the inverse $|X|^{-1}$).
Transforming the index $k$ on the r.h.s.\ onto a scalar superfield according to  (\ref{mes1}) and using the
projection operators (\ref{projectors2}) we can write down the general solution to (\ref{Asol}).
Consequently, the enhanced susy transformation of the fermionic connection is of
the form:
\begin{align}
  \label{dA}
  \delta {\ca A}_{\alpha\,ij}{}^a{}_d &= \epsilon^\beta\left(\,\varepsilon_{ijmn}\,\bar\del^{m\,a}|X|\,
      \bar G^{\, n} _{\alpha\beta\,d} +\ca{P}^a{}_{c}\, H_{\alpha\beta\,ij}{}^c{}_d  \,\right)\nonumber\\[5pt]
      &\hspace{1.5cm} +\fr{1}{2}\epsilon_{ijkl}\,\bar\epsilon^\beta
   \left(\,\varepsilon^{klmn}\del_{m\,d}|X| G^{\ \ a}_{\alpha\beta\, n} +
          \ca{P}^c{}_d\, \bar H^{kl}_{\alpha\beta}{}^a{}_c \,\right) \ ,
\end{align}
where following our general conventions $\bar G^{\, n} _{\alpha\beta\,d} = (G^{\ \ d}_{\alpha\beta\, n})^\ast$
and $H_{\alpha\beta\,ij}{}^c{}_d$  is a (composite) superfield in the ${\bf \bar 6}$.

With this general solution to the algebraic part of the invariance condition (\ref{dmatter2}) for the 
matter constraint one obtains the following constraint for the monopole operator superfield:
\begin{align}
  \label{dM}
 \nabla_{\alpha\,ij}\ca{M}^{ab}\,\bar{\Psi}_{\beta\,b}^{k }|_{\bf \overline{20}}&= \left(
   \varepsilon_{ijmn}\,\bar\del^{m\,a}|X|\,\bar G^{\, n} _{\alpha\beta\,d}\Phi^{k\,d}
  + \ca{P}^{a}{}_c \, H_{\alpha\beta\,ij}{}^c{}_d \Phi^{k\,d}\right.\nonumber\\[2pt]
   &\hspace{4cm}\left. +\ \varepsilon_{\alpha\beta}(\ca{M}f)^{a,ce}{}_d\,\bar\Phi_{i\,c}\bar\Phi_{j\,e}\Phi^{k\,d}
   \right)|_{\bf \overline{20}}\ ,
\end{align}
which defines the super-derivative monopole operator field (though contracted with the fermionic superfield)
up to a contribution in the ${\bf \bar 4}$, in terms of the yet unconstrained composite superfields 
$G$ and $H$. The unspecified ${\bf \bar 4}$ contribution
combines with all the other  ${\bf \bar 4}$ contributions which were not written explicitly to the
enhanced susy variation of the fermionic superfield, see (\ref{dc2}). 

A few comments are in order:\\
{\em i.)} 
Contrary to the previous case with a covariantly constant monopole operator now there is no 
algebraic condition on the monopole operator at this stage. One might think that the more general procedure
presented here would also help for the case of a covariantly constant monopole operator. In this case, 
equation (\ref{dM}) imposes an algebraic condition on the superfields $G$, $H$ and $\ca{M}$ which 
has no other obvious solution then the one considered in the previous subsection. In particular the 
invariance of the gauge constraint cause further problems.    \\
{\em ii.)} The composite superfields 
$G$ and $H$ are not further specified yet. The requirement that the also the gauge field constraint
(\ref{defW}) is invariant under the enhanced susy will determine also the super-derivatives of these fields,
and thus in combination with (\ref{dM}) form a system of superspace constraints for the monopole operator
 superfield. 
This is what we consider next. \\
{\em iii.)} To obtain the super-derivative for the monopole operator without being contracted 
with the fermionic field one has to factor out a fermionic superfield in the whole equation. This can be done
in an obvious way for the composite fields $G$ and $H$ but not for the monopole operator field itself without any 
further assumption/derivation of the composite structure of the monopole operator. We leave this question
for a followup investigation, where we analyze the complete monopole operator constraint system, which 
we develop here.
\\

\noindent
{\bf Gauge Field Constraint.} To complete the invariance of our original 
constraint system, (\ref{dc2}), (\ref{defW}), the  enhanced susy variations determined so far have to 
satisfy the invariance condition (\ref{dgauge}). In deriving the susy variation of
the fermionic connection (\ref{dA}) we demonstrated explicitly the principle methods 
which are needed. The derivations of the following expressions, though more lengthy, work out in a
similar way. All needed identities involving the above introduced mesonic operators are given in appendix 
\ref{Apmes}.

Due to the properties under complex conjugation it is sufficient to consider the
part of the invariance condition  (\ref{dgauge}) proportional to  $\epsilon$, the 
$\bar\epsilon$-part is then automatically satisfied. 
Writing the enhanced susy variation of the bosonic connection as
\begin{equation}
  \label{dAb1}
  \delta{\cal A}_{\alpha\beta}{}^a{}_b:=\epsilon^{\gamma}B_{\alpha\beta,\gamma}{}^a{}_b
                   -\bar\epsilon^{\gamma}(B_{\alpha\beta,\gamma}{}^b{}_a)^\ast\ ,
\end{equation}
so that it respects the reality condition (\ref{eq:1.3}) and $B_{\alpha\beta,\gamma} = B_{(\alpha\beta),\gamma}$, 
the $\epsilon$-part of the invariance condition
(\ref{dgauge}) writes as
\begin{align}
  \label{dgauge2}
  \nabla_{\alpha\,ij}&\delta_{\epsilon}{\cal A}_{\beta\,kl}{}^a{}_b
   +\nabla_{\beta\,kl}\delta_{\epsilon}{\cal A}_{\alpha\,ij}{}^a{}_b =\nonumber\\[3pt]
  &\hspace{1cm} i\,\epsilon^{\gamma}\{\varepsilon_{ijkl}B_{\alpha\beta,\gamma}{}^a{}_b
     +\fr{1}{2}\epsilon_{\alpha\beta}(\ca{M}f)^{d,ca}{}_d\,
      (\varepsilon_{m,ij[k}\bar\Phi_{l]c} - \varepsilon_{m,kl[i}\bar\Phi_{j]c})\,\bar\Psi^m_{\gamma\,d}\}\ ,
\end{align}
where $\delta_{\epsilon}{\cal A}_{\alpha\,ij}{}^a{}_b$ is the $\epsilon$-part in the enhanced susy variation given
in (\ref{dA}). Explicitly, in terms of the composite superfields $G$ and $H$ one finds
\begin{align}
  \label{full}
 &\nabla_{\alpha\,ij}\delta_{\epsilon}{\cal A}_{\beta\,kl}{}^a{}_b
   +\nabla_{\beta\,kl}\delta_{\epsilon}{\cal A}_{\alpha\,ij}{}^a{}_b =\nonumber\\[5pt]
 &\hspace{0.5cm} -\epsilon^{\gamma}\{
      \left(\varepsilon_{ijmn}\bar\del^{m\,a}|X|\nabla_{\beta\,kl}\bar G^{\, n} _{\alpha\gamma\,b}
       -i\, \varepsilon_{ijmn}\bar G^{\, n} _{\alpha\gamma\,b}\Psi^c_{\beta[k}\del_{l]c}\bar\del^{m\,a}|X|
       + i\, \nabla_{\alpha\,ij}(\ca{P}^a{}_c\,H_{\beta\gamma\,kl}{}^c{}_b) \right. \nonumber\\[3pt] 
     &\hspace{1.2cm} \left.  + \textstyle{\binom{\alpha}{ij}\leftrightarrow \binom{\beta}{kl}}
      \right)
       -\fr{i}{2}\,\varepsilon_{ijmn}\,\varepsilon_{klrs}\,
          (\bar G^{\, n} _{\alpha\gamma\,b}\bar\Psi^r_{\beta\,c}-\bar G^{\, r} _{\beta\gamma\,b}\bar\Psi^n_{\alpha\,c})\,
         \bar\del^{s\,c}\bar\del^{m\,a}|X|\}\ .
\end{align}
The invariance condition (\ref{dgauge2}), with the l.h.s. given by (\ref{full}), has to be satisfied 
so that also the gauge field constraint is invariant under the enhanced supersymmetry. This
defines superspace constraints for the composite fields $G$ and $H$ which together with 
(\ref{dM}) form a constraint system for the monopole operator superfield $\ca{M}^{ab}$. We now
extract the constraints for the composite superfields $G$ and $H$.

Contracting (\ref{full}) with a scalar superfield $\bar\Phi_{p\,a}$ one obtains an equation which
contains only the super-derivative of the $G$-field, without any (non-invertible) field dependent pre-factor. 
To achieve this one uses $\bar\Phi_{p\,a}\bar\del^{m\,a}|X|=\delta^m{}_p|X|$, see appendix \ref{Apmes}, and 
further, that with (\ref{projectors2}) and the matter constraint (\ref{dc2})
\begin{equation}
  \label{eq:1}
   \bar\Phi_{p\,a}\nabla_{\alpha\,ij}(\ca{P}^a{}_c\,H_{\beta\gamma\,kl}{}^c{}_b)=
   -\fr{i}{2}\,\varepsilon_{ijpq}\bar\Psi^q_{\alpha\,a}(\ca{P}^a{}_c\,H_{\beta\gamma\,kl}{}^c{}_b)\ .
\end{equation}
The resulting equation has a unique solution for the super-derivative of the $G$-field, which is given 
by
\begin{align}
  \label{dG}
   \nabla_{\alpha\,ij}\bar G^{\, n} _{\beta\gamma\,b}=&\, \fr{1}{|X|}\left\{
   2\,i\,\delta^n{}_{[i}\bar\Phi_{j]a}\,B_{\alpha\beta,\gamma}{}^a{}_b
     +\fr{i}{2}\varepsilon_{\alpha\beta}(\ca{M}f)^{d,ce}{}_b\,
          (\delta^n{}_{[i}\bar\Phi_{j]c}\bar\Phi_{m\,e}\bar\Psi^m_{\gamma\,d}
           +\bar\Phi_{ic}\bar\Phi_{je}\bar\Psi^n_{\gamma\,d})\right.\nonumber\\[3pt]
   & + i\,\bar G^{\, n} _{\beta\gamma\,b}\Psi^c_{\alpha[i}\,\del_{j]c}|X|
      +\fr{i}{2}\,\bar\Psi^n_{\beta\,c}\,\ca{P}^c{}_aH_{\alpha\gamma\,ij}{}^a{}_b\nonumber\\[3pt]
      &\left.\hspace{4cm}+\,\fr{i}{2}\,\varepsilon_{ijrs}\,
    (\bar G^{\, n} _{\beta\gamma\,b}\bar\Psi^r_{\alpha\,c} 
                       - \bar G^{\, r} _{\alpha\gamma\,b}\bar\Psi^n_{\beta\,c})\,\bar\del^{s\,c}|X|\right\}\ .
\end{align}
Inserting  this solution back into the original invariance condition (\ref{dgauge2}) 
gives a superspace constraint equation for the $H$-field:
\begin{align}
  \label{dH}
  \ca{P}^e{}_a\nabla_{\alpha\,ij}&H_{\beta\gamma\,kl}{}^a{}_b|_{({\bf 3},{\bf 1}\oplus{\bf 20})\oplus ({\bf 1},{\bf 15})} =
              \nonumber\\[4pt]
    &\hspace{0.0cm} \ca{P}^e{}_a 
    \{ i\,\varepsilon_{ijmn}\,\bar G^{\, n} _{\alpha\gamma\,b}\Psi^c_{\beta[k}\,\del_{l]c}\,\bar\del^{m\,a}|X|
      -\fr{1}{|X|}\,\Psi^a_{\alpha[i}\,\del_{j]c}|X|\, H_{\beta\gamma\,kl}{}^c{}_b\nonumber\\[3pt]
   & \hspace{0.8cm}\left.  -\fr{i}{2}\left(\varepsilon_{ijkl}\,B_{\alpha\beta,\gamma}{}^a{}_b
       +\varepsilon_{\alpha\beta} 
    (\ca{M}f)^{d,ca}{}_b\, \varepsilon_{mij[k}\bar\Phi_{l]c}\bar\Psi^m_{\gamma\,d} \right)\right\}
         |_{({\bf 3},{\bf 1}\oplus{\bf 20})\oplus ({\bf 1},{\bf 15})},
\end{align}
which fixes the ${({\bf 3},{\bf 1}\oplus{\bf 20})\oplus ({\bf 1},{\bf 15})}$ content  
of the super-derivative of the $H$-field, but only the projection onto the eigenspace of
the projector $\ca{P}^e{}_a$ defined in (\ref{projectors}).

The  constraint equations for the $G$- and $H$- field (\ref{dG}), (\ref{dH})
together with the constraint equation for the monopole operator $\ca{M}$ (\ref{dM})  define
the constraint system for the monopole operator superfield. This constraint system
was derived starting from a single assumption, namely the transformation (\ref{dbos}),
and the requirement that this transformation is part of the enhanced supersymmetry
implemented as an internal fermionic ($\ca{N}=2$ super) symmetry of our matter-gauge 
constraint system (\ref{dc2}), (\ref{defW}). 

At this stage we did not encounter any algebraic condition on the monopole operator field $\ca{M}$, 
as it is the case for a covariantly constant monopole operator and 
which allowed only for the gauge group $U(2)\times U(2)$. Actually, we did not encounter
any condition on the gauge group yet, so in principle all gauge groups of the classification
below (\ref{fsol}) are allowed. We expect that a further study of the monopole constraint system
will select the gauge group to be $U(N)\times U(N)$. The allowed CS-levels are in principle
given by the general analysis following (\ref{weights}). To see this condition explicitly 
one might have to consider the operators, at least in principle, as quantum fields,
as for example in the case for the quantization condition of the CS-level itself. 
See also the discussion below (\ref{u2u2mop}).
 In this regard we just want to mention that in the equations of the 
constraint system for the monopole operator superfield, i.e.\ (\ref{dG}), (\ref{dH}) and  
(\ref{dM}), an (abelian)\footnote{Or a covariantly constant operator living in the eigenspace of the 
projector $\ca{P}$} covariantly constant operator can be factored out.
\\

\noindent
A number of things remains to be done. First one has to analyze the monopole superfield
constraint system (\ref{dG}),(\ref{dH}), (\ref{dM}) analogously to the procedure 
applied to the matter-gauge constraint system as done in sections \ref{sec:msf} and 
\ref{sec3} and as it was presented in greater detail in \cite{Samtleben:2009ts}
for the $\ca{N}=8$ case. One has to see if it will be necessary to strip off the 
fermionic field in equ. (\ref{dM}) and if so, if the analysis of the constraint system
provides enough information about the composite structure of the monopole operator field
$\ca{M}$ to do so or if further assumptions are necessary.

Second, we did not write down explicitly the enhanced susy transformation for the fermionic 
superfield, which is given by all the ${\bf \bar 4}$ contributions in the invariance condition
of the matter constraint (\ref{dmatter}). Given the transformation rule
for the fermionic superfield one can check the algebra of the enhanced symmetry 
transformations which might give additional information to specify the 
monopole operator superfield further.

We leave these points for a follow up investigation to the  structure developed here. Eventually,
a detailed analysis of the here developed constraint system will lead also to
space-time e.o.m.\ for the composite monopole superfield,  as we obtained the 
superfield e.o.m (\ref{confeom}) from the matter- and gauge constraint (\ref{dc2}), (\ref{defW}).
These space-time e.o.m.\ for the composite monopole superfield might then describe the 
dynamics of a theory dual to the ABJM model, in the sense of the three-dimensional 
mirror symmetry \cite{Intriligator:1996ex}. This would finally be  a nonabelian gauge theory analogon of the explicit 
duality relations for
 two-dimensional soliton models \cite{Coleman:1974bu,Mandelstam:1975hb}. We will address
these issues in  future work.

 \bigskip

\noindent
{\bf Acknowledgements:}
We thank F.~Delduc and K.~Gawedzki for various helpful discussions. 
This work is supported in part by the Agence Nationale de la Recherche (ANR).


\appendix

\section{$SO(6) \sim SU(4)$ conventions}

\subsection*{$SO(6)$ $\Gamma$-matrices}

Dirac spinors of $SO(6)$ have eight components. The irreducible representations are  given 
by four component Weyl spinors (though $SO(6)$ provides also Majorana spinors). The Weyl representation
of the hermitian $SO(6)$ $\Gamma$-matrices  is of the form
\begin{equation}
 \label{eq:a11.1}
 \hat{\Gamma}^I=\left [\begin{array}{cc}0&\Gamma^I\\\bar{\Gamma}^I&0\end{array}\right]\ ,
\end{equation} 
where hermiticity implies $\bar{\Gamma}^I = (\Gamma^I)^\dagger$ and these matrices 
satisfy $\Gamma^{(I}\bar\Gamma^{J)}=\delta^{IJ}\unit$. We denote the components of these matrices  by
\begin{equation}
  \label{eq:a11.2}
      \Gamma^{Iij} \quad , \quad \bar{\Gamma}^I_{ij}\quad \mathrm{with}\quad i,j=1,\ldots,4\ .   
\end{equation}
A particular representation which makes the relation to $SU(4)$ transparent is given by \cite{Terashima:2008sy}
\csep
\begin{eqnarray}
  \label{eq:a11.3}
   \Gamma^1=\sigma_2\otimes\unit_2 \quad &\Gamma^2=-i\sigma_2\otimes\sigma_3 &
              \quad\Gamma^3=i\sigma_2\otimes\sigma_1 
\nonumber\\
  \quad\Gamma^4=-\sigma_1\otimes\sigma_2 \quad&\!\!\!\!\!\!\!\Gamma^5=\sigma_3\otimes\sigma_2 &
       \quad\Gamma^6=-i\unit_2\otimes\sigma_2
\end{eqnarray}
These matrices are anti-symmetric, i.e.\ $\Gamma^{I ij}=-\Gamma^{Iji}$ and satisfy the reality condition
\begin{equation}
  \label{eq:a11.4}
  (\Gamma^{Iij})^\ast=\fr{1}{2}\epsilon_{ijkl}\Gamma^{I kl}=-\bar{\Gamma}^I_{ij}\ ,
\end{equation}
further, the contractions satisfy
\begin{equation}
  \label{eq:a11.5}
  \Gamma_I^{ij}\bar\Gamma^I_{kl}=-4\delta^{ij}_{kl}\quad , \quad \Gamma_I^{ij}\Gamma^{I kl}=2\epsilon^{ijkl}\ ,
\end{equation}
where $\delta^{i_1\ldots i_n}_{k_1\ldots k_n}=\delta^{[i_1}_{k_1}\ldots\delta^{i_n]}_{k_n}$ and the 
totally antisymmetric epsilon symbol
is defined as $\epsilon^{1234}=\epsilon_{1234}=1$ so that
\begin{equation}
  \label{eq:a11.6}
   \epsilon^{i_1\ldots i_n j_1\ldots j_{D-n}}\epsilon_{i_1\ldots i_n k_1\ldots k_{D-n}} =
   (D-n)!\ n!\  \delta^{j_1\ldots j_{D-n}}_{k_1\ldots k_{D-n}}\ ,
\end{equation}
with $D=4$ in this case. This is all that wee need of $\Gamma$-matrix relations to show how $SO(6)$ 
representations are related to $SU(4)$ representations. The main text is formulated in terms
of $SU(4)$ representations thus avoiding any explicit $\Gamma$-matrix relations.

\subsection*{$SU(4)$ representations}\label{A12}
The $SO(6)$ generators write with (\ref{eq:a11.1}) as
\begin{equation}
  \label{eq:a12.1}
\hat\Sigma^{IJ}:=\fr{1}{2}[\hat\Gamma^I,\hat\Gamma^J]=
     \left [\begin{array}{cc}\Gamma^{IJ}&0\\0&\bar{\Gamma}^{IJ}\end{array}\right]\ ,
\end{equation}
where the irreducible 4 times 4 blocks
\begin{eqnarray}
  \label{eq:a12.2}
   \Gamma^{IJ i}{}_j:=\Gamma^{[I ik}\bar\Gamma^{J]}_{kj} \quad ,\quad 
    \bar\Gamma^{IJ}{}_i{}^j:=\bar\Gamma^{[I}_{ik}\Gamma^{J]kj}, 
\end{eqnarray}
satisfy $\mathrm{tr}\,\Gamma^{IJ}=0$ and
$(\Gamma^{IJ})^\dagger=-\Gamma^{IJ}=\Gamma^{JI}$, with the same relations for $\bar\Gamma^{IJ}$. 
These matrices are therefore the 15 anti-hermitian generators of $SU(4)$ and $SU(\bar 4)$ transformations,
 respectively (the $SU(4)$ algebra is implied by the $SO(6)$ algebra). The above given 
relations for $\Gamma^I$ and $\bar\Gamma^I$ imply the conjugation properties
\begin{equation}
  \label{eq:a12.3}
(\Gamma^{IJ i}{}_j)^\ast=\bar\Gamma^{IJ}{}_i{}^j=-\Gamma^{IJ j}{}_i\ .  
\end{equation}

We give here the notation for some $SU(4)$ representations which occur frequently in the main text but
without taking the gauge group structure into account. Modifications 
for fields in the adjoint representation of the gauge group are given at the end of  appendix \ref{B}.             

\underline{Representations ${\bf 4}$ and $\bar {\bf 4}$}: According to the conventions as given in (\ref{eq:a12.2}) 
a field in the ${\bf 4}$ carries an upper index, e.g.\ $\Phi^i$, and a field in the  $\bar {\bf 4}$ carries
a lower index, e.g.\ $\Psi_i$. Complex conjugation maps one representation into the other, see (\ref{eq:a12.3}), 
which we denote by
\begin{equation}
  \label{eq:a12.4}
(\Phi^i)^\ast=:\bar\Phi_i \quad , \quad   (\Psi_i)^\ast=:\bar\Psi^i\ .
\end{equation}

\underline{Representations  ${\bf 6}$, $\bar {\bf 6}$ and their real form}: The ${\bf 6}$ appears in the 
product ${\bf 4}\otimes{\bf 4}$ (the $\bar {\bf 6}$ in the complex conjugated thereof)  and
it is therefore natural to use the conventions
\begin{equation}
  \label{eq:a12.4.1}
     {\bf 6}: v^{ij}\quad\mathrm{with}\quad v^{ij}=v^{[ij]}\quad , \quad
     \bar{\bf 6}: u_{ij}\quad\mathrm{with}\quad  u_{ij}=u_{[ij]} \ .
\end{equation}
These are $6$ dimensional complex representations. Complex conjugation relates this representation 
to each other and
thus we have the convention $(v^{ij})^\ast=:\bar v_{ij}$ and accordingly for $u_{ij}$.  $SO(6)$ has a real
representation ${\bf 6}$ which translates into a real ${\bf 6}$ or ${\bf \bar 6}$ of $SU(4)$ by
\begin{equation}
  \label{eq:a12.4.2}
  v^{ij}:=\fr{1}{2} \Gamma^{I\ ij}v_I\quad \mathrm{or}\quad 
 v_{ij}:=- \fr{1}{2} \bar\Gamma^I{}_{ij}v_I=\fr{1}{2}\epsilon_{ijkl}v^{kl}\ ,
\end{equation}
where we have used (\ref{eq:a11.4}). This implies  the reality condition
\begin{equation}
  \label{eq:a12.4.3}
  {\bar v}_{ij}:=(v^{ij})^\ast= \fr{1}{2}\epsilon_{ijkl}v^{kl}=v_{ij}\ .
\end{equation}
Since $\epsilon_{ijkl}$ is an invariant 
$SU(4)$ tensor this reality condition is invariant. In fact it is the natural condition to impose 
on representations ${\bf 6}$ of $SU(4)$. Due to the identification (\ref{eq:a12.4.2}), (\ref{eq:a12.4.3}) 
of the real forms of the ${\bf 6}$ and ${\bf \bar 6}$  it is unnecessary to 
differentiate between them and we will generally speak if the real ${\bf 6}$ in either case.

\underline{Representations  ${\bf 15}$, ${\bf \overline{15}}$ and their real form}: The  ${\bf 15}$ appears in
the product ${\bf 4}\otimes\bar {\bf 4}$ (the $ {\bf  \overline{15}}$ in the complex conjugated thereof) and
it is therefore natural to use the conventions
\begin{equation}
  \label{eq:a12.5}
     {\bf 15}: W^i{}_j\quad\mathrm{with}\quad W^i{}_i=0\quad , \quad
     {\bf \overline{15}}: U_i{}^j\quad\mathrm{with}\quad U_i{}^i=0\ .
\end{equation}
These are $4\times4-1=15$ dimensional complex representations. Complex conjugation we denote according to
(\ref{eq:a12.3}) by $(W^i{}_j)^\ast=:\bar W_i{}^j$ and similar for $U_i{}^j$. $SO(6)$ has also a 
representation ${\bf 15}$
which is real, $W_{IJ}=W_{[IJ]}$. This is translated into a real ${\bf 15}$ or ${\bf  \overline{15}}$ of $SU(4)$  by
\begin{equation}
  \label{eq:a12.6}
   W^i{}_j:=-\fr{1}{2}\Gamma^{IJ i}{}_jW_{IJ}\quad\mathrm{or}\quad 
       W_i{}^j= -\fr{1}{2}\bar\Gamma^{IJ}{}_i{}^j W_{IJ}.
\end{equation}
This then implies with (\ref{eq:a12.3}) the reality condition
\begin{equation}
  \label{eq:2}
  \overline{W}_i{}^j:=(W^i{}_j)^\ast=-W^j{}_i=W_i{}^j\ .
\end{equation}
It is easy to see that this reality condition is invariant under $SU(4)$ transformations. Again there is no reason 
to differentiate between this two real forms and we will express everything in terms of the real form of the ${\bf 15}$.

\section{Superspace and Connections}\label{B}

The $\ca{N}=6$ susy algebra without (non)-central charges is given by six 3D-Majorana spinors 
transforming in the ${\bf 6 }$ of $SO(6)$ and satisfying the algebra\footnote{$SO(6)$ vector indices $I,\ldots$ 
are raised and lowered with the Kronecker delta and therefore one does not have to pay any attention to their position.}
\begin{equation}
  \label{eq:a13.1}
  \{Q^I_{\alpha},Q^J_{\beta}\}=2\delta^{IJ}P_{\alpha\beta}\ .
\end{equation}
Accordingly, the susy parameters $\epsilon^{I\alpha}$ and the superspace coordinates $\theta^{I\alpha}$
are also given by 3D-Majorana spinors in the ${\bf 6}$ ($I=1,\ldots,6$). Our 3D conventions are such that
Majorana spinors are  real (see appendix \ref{so3}). 
To represent the the algebra (\ref{eq:a13.1}) and to form susy covariant expressions 
one introduces the following differential operator,
\begin{equation}
  \label{eq:a13.2}
  D_{\alpha\, I}:=\del_{\alpha\, I}+i \theta^\beta_I\del_{\alpha\beta}\quad , \quad 
   Q_{\alpha\, I}:=\del_{\alpha\, I}-i \theta^\beta_I\del_{\alpha\beta}\ ,
\end{equation}
such that $\{ D_{\alpha\, I},Q_{\beta\, J}\}=0$ and
\begin{equation}
  \label{eq:a13.3}
  \{ Q_{\alpha\, I},Q_{\beta\, J}\}=-\{ D_{\alpha\, I},D_{\beta\, J}\}=-2i \delta_{IJ}\del_{\alpha\beta} \ ,
\end{equation}
where $\del_{\alpha I}\theta^{\beta J}=\delta^\alpha_\beta \delta^J_I$. These operators are hermitian in the 
Hilbert space of superfields and have the following definite properties under complex conjugation 
(therefore the $i$ in the definition):
\begin{equation}
  \label{eq:a13.4}
  (D_{\alpha\, I} X)^\ast=(-)^{|X|+1}D_{\alpha\, I}\bar X \quad , \quad (\del_{\alpha\beta})^\ast=\del_{\alpha\beta}\ ,
\end{equation}
where $|X|$ is the fermion parity of a superfield $X$ ($=0,1$ for bosonic/fermionic). 

We now write the ${\bf 6}$ of $SO(6)$ in these definitions as real ${\bf 6}$ of $SU(4)$ according
to (\ref{eq:a12.4.2}):
\begin{equation}
  \label{eq:13.5}
  \theta^{\alpha\, ij}=\fr{1}{2}\Gamma_I^{ij}\theta^{\alpha I} \quad , \quad 
   D_{\alpha\, ij}=-\fr{1}{2} \bar \Gamma^I_{ij}D_{\alpha I} \ .
\end{equation}
With this definitions $\theta^{\alpha\, ij}$ satisfies the conjugation property (\ref{eq:a12.4.3}). The 
relations (\ref{eq:a12.3}) and (\ref{eq:a13.4}) imply for the superderivative
\begin{equation}
  \label{eq:a13.6}
   (D_{\alpha\, ij} X)^\ast=(-)^{|X|+1}\ \fr{1}{2}\epsilon^{ijkl}D_{\alpha\, kl}\bar X
    =:(-)^{|X|+1}D_\alpha^{ij}\bar X\ .
\end{equation}
For the fermionic derivative one has now 
$\del_{\alpha\, ij}\theta^\beta_{kl}=\fr{1}{2}\epsilon_{ijkl}\delta^\beta_\alpha$ and the anti-commutator
(\ref{eq:a13.3}) writes as
\begin{equation}
  \label{eq:a13.7}
  \{D_{\alpha\, ij},D_{\beta\, kl}\}=-\{Q_{\alpha\, ij},Q_{\beta\, kl}\}=i \epsilon_{ijkl}\del_{\alpha\beta}\ .
\end{equation}

{\bf Covariant derivatives.} We introduce connections on superspace in the following way
\begin{eqnarray}
  \label{eq:a13.8}
  \nabla_{\alpha\beta}:=\del_{\alpha\beta}+\ca{A}_{\alpha\beta}\quad &\mathrm{with}&\quad 
            \ca{A}_{\alpha\beta}= \ca{A}_{\alpha\beta}^M T_M \nonumber\\
  \nabla_{\alpha\, I}:=D_{\alpha\, I}+\ca{A}_{\alpha\, I}\quad &\mathrm{with}&\quad
               \ca{A}_{\alpha\, I}= \ca{A}_{\alpha\, I}^M i T_M\ ,
\end{eqnarray}
with anti-hermitian generators $(T_M)^\dagger=-T_M$. With these definitions and taking the 
coefficient fields $\ca{A}_{\alpha\beta}^M$, $\ca{A}_{\alpha\, I}^M$ to be real superfields,
the connection one-forms are (anti)hermitian, i.e.\ $(\ca{A}_{\alpha\beta})^\dagger=-\ca{A}_{\alpha\beta}$
and $(\ca{A}_{\alpha\, I})^\dagger=\ca{A}_{\alpha\, I}$ and so are the resulting field strengths
$(\ca{F}_{\alpha\beta,\gamma\delta})^\dagger= -\ca{F}_{\alpha\beta,\gamma\delta}$ and 
$(\ca{F}_{\alpha\,I,\beta\,J})^\dagger= \ca{F}_{\alpha\,I,\beta\,J}$. The ${\bf 6}$ of $SO(6)$ we again write 
as real ${\bf 6}$ or ${\bf \bar 6}$ of $SU(4)$ according to  (\ref{eq:a12.4.2}) as
\begin{equation}
  \label{eq:a13.9}
  \nabla_{\alpha\, ij}:=-\fr{1}{2}\bar\Gamma^I_{ij}\nabla_{\alpha\,I}\quad , \quad
     \nabla_{\alpha}^{ij}:=\fr{1}{2}\Gamma^{I\,ij}\nabla_{\alpha\,I}\ .
\end{equation}

For the action of the generators of the gauge group in a representation $R$ and $\bar R$, respectively, 
we introduce the following convention:
\begin{equation}
  \label{eq:a13.10}  
T_M\cdot X^a:=T_M{}^a{}_b X^b \quad , \quad T_M\cdot Y_b:=-T_M{}^a{}_b Y_a \ , 
\end{equation}
where we denote a field in the representation $R$/$\bar R$ with an upper/lower index from the range 
$a,b,\ldots$. The action on $\bar R$ follows from the action on $R$ by complex conjugation
where $(X^a)^*=:\bar X_a$. The action on tensor 
products of these representations is defined accordingly. The covariant derivatives of a field in the $R$ and
its complex conjugate thus write as (suppressing space-time and $SU(4)$ indices)
\begin{eqnarray}
  \label{eq:a13.11}
  \nabla\Phi^a&=&D\Phi^a+\ca{A}\cdot\Phi^a=D\Phi^a+\ca{A}^a{}_b\Phi^b \ , \nonumber\\ 
      \nabla\bar\Phi_b&=&D\bar\Phi_b+\ca{A}\cdot\bar\Phi_b=D\bar\Phi_b-\ca{A}^a{}_b\bar\Phi_a\ ,
\end{eqnarray}
where $D$ stands either for $D_{\alpha\,ij}$ or $\del_{\alpha\beta}$.

With the definitions (\ref{eq:a13.9}) we thus obtain the following properties under complex conjugation:
\begin{eqnarray}
  \label{eq:a13.12}
  (\nabla_{\alpha\beta}X^a)^\ast&=&\nabla_{\alpha\beta}\bar X_a=
                \del_{\alpha\beta}\bar X_a-\ca{A}_{\alpha\beta}{}^b{}_a\bar X_b\\
  (\nabla_{\alpha\,ij}X^a)^\ast&=&
       (-)^{|X|+1}\fr{1}{2}\ \epsilon^{ijkl}\nabla_{\alpha\,kl}\bar X_a =
        (-)^{|X|+1}\nabla_\alpha^{ij}\bar X_a \ .
\end{eqnarray}
The fermionic field strength in $SU(4)$ notation has therefore the conjugation property
 $(\ca{F}_{\alpha\beta,\gamma\,ij})^\dagger=
\fr{1}{2}\epsilon^{ijkl}\ca{F}_{\alpha\beta,\gamma\,kl}$.

{\bf Gauge field constraint.} Analogous to the $\ca{N}=8$ case \cite{Samtleben:2009ts} we impose 
the constraint
\begin{equation}
  \label{eq:a13.13}
  \{\nabla_{\alpha\,I},\nabla_{\beta\,J}\}=2 i (\delta_{IJ}\nabla_{\alpha\beta}+\varepsilon_{\alpha\beta}W_{IJ})\ .
\end{equation}
Translating this into $SU(4)$ representations we multiply this equation with 
$(-\hal\bar\Gamma^I_{ij})(-\hal\bar\Gamma^J_{kl})$. Using above relations one has
\begin{equation}
  \label{eq:a13.14}
  \bar\Gamma^I_{ij}\bar\Gamma^J_{kl}W_{IJ}=2\epsilon_{mij[k}W^m{}_{l]}=-2\epsilon_{mkl[i}W^m{}_{j]}\ ,
\end{equation}
where we chose the real ${\bf 15}$ of $SU(4)$, a similar relation holds for the real ${\bf  \overline{15}}$. 
The constraint (\ref{eq:a13.13}) thus writes as
\begin{equation}
  \label{eq:a13.15}
   \{\nabla_{\alpha\,ij},\nabla_{\beta\,kl}\} =  i(\epsilon_{ijkl}\nabla_{\alpha\beta}+
      \varepsilon_{\alpha\beta}\epsilon_{mij[k}W^m{}_{l]})
\end{equation}

Fields in the gauge sector like $W^i{}_j$ and $\lambda_{\alpha\,ij}$, $\rho_{\alpha\,ij}$ etc. (see the main text)
live in the Lie algebra $\mathfrak{g}$. We expand bosonic 
superfields like $W^i{}_j$ as $W^i{}_j=W^M{}^i{}_jT_M$ and fermionic ones with an extra $i$, i.e.\ 
as $\lambda_{\alpha\,ij}=\lambda^M_{\alpha\,ij}iT_M$ etc.,  with anti-hermitian generators 
$(T_M)^\dagger=-T_M$. This means in replacing complex conjugation by hermitian conjugation
the conditions due to $SU(4)$ representations as given in appendix \ref{A12} are unchanged for fermionic fields and 
receive an extra minus for bosonic fields. For example
\begin{equation}
  \label{eq:a13.16}
  (W^i{}_j)^\dagger=W^j{}_i\quad \textrm{and}\quad 
         (\lambda_{\alpha\,ij})^\dagger = \fr{1}{2}\epsilon^{ijkl}\lambda_{\alpha\,kl}\ ,
\end{equation}
for a bosonic field in the real ${\bf 15}$ and a fermionic field in the real ${\bf 6}$.
  For conformal theories the potential $W^i{}_j$ is explicitly given in (\ref{Wconf}). For this potential
the derived composite fields in the gauge sector, see section \ref{gaugesector}, are obtained 
according to (\ref{compf}) as follows:
\begin{align}
  \label{compconf}
 &( \lambda_{\alpha\,ij})^a{}_b=i \F{a}{c}{b}{d}\left(\Psi^d_{\alpha\,[i}\bar\Phi_{j]\,c}
     +\hal\epsilon_{ijkl}\Phi^{k\,d}\bar\Psi^l_{\alpha\,c}\right) \, ,\ \ 
  (\rho_{\alpha\,ij})^a{}_b =i \F{a}{c}{b}{d}\ \bar\Phi_{c\,(i}\Psi^d_{j)\,\alpha}\, ,\nonumber\\[3pt]
 & (V^i{}_j)^a{}_b=\fr{i}{4} \vep^{\alpha\beta}\F{a}{c}{b}{d}\left(\Psi^d_{\alpha\,j}\bar\Psi^i_{\beta\,c}
          -\fr{1}{4}\delta^i{}_j(\Psi_{\alpha}\bar\Psi_\beta)^d{}_c\right)
          -\fr{1}{4}\F{g}{c}{h}{d}\F{d}{a}{e}{b}\left(\Phi^{i\,h}\bar\Phi_{j\,g}(\Phi\bar\Phi)^e{}_c \right) \nonumber \\[3pt]
  &\hspace{20mm}    -\fr{1}{4}\F{a}{c}{b}{d}\F{d}{g}{e}{h}\left(\Phi^{i\,h}\bar\Phi_{j\,g}(\Phi\bar\Phi)^e{}_c
                  -\hal\delta^i{}_j(\Phi\bar\Phi)^h{}_g (\Phi\bar\Phi)^e{}_c\right)\ ,
\end{align}
where $(\Phi\bar\Phi)^d{}_c=\Phi^{d\,k}\bar\Phi_{k\,c}$ etc. is short for contracted $SU(4)$ indices.

\section{Mesonic Operators}\label{Apmes}

We develop and collect here in some detail the structure of mesonic operators, in particular the 
$SU(4)$ determinant thereof, which were introduced in the main text in (\ref{mesop}). 
We define the mesonic operators as the gauge invariant quantities
\begin{equation}
  \label{mesa1}
  X^i{}_j:=\Phi^{i\,a}\bar\Phi_{j\,a}\ \ \Rightarrow \ \ (X^i{}_j)^\ast = X^j{}_i \ ,
\end{equation}
which transform in the $({\bf 4},{\bf \bar 4})$ under $SU(4)$. 
With this definition the determinant of the mesonic operators can be written as
\begin{equation}
  \label{mesa2}
  \mathrm{det}(X^i{}_j) =:|X|=\fr{1}{24}\,\varepsilon_{ijkl}(\Phi^{i\,a}\Phi^{j\,b}\Phi^{k\,c}\Phi^{l\,d})
      \, \varepsilon^{mnpq}(\bar\Phi_{m\,a}\bar\Phi_{n\,b}\bar\Phi_{p\,c}\bar\Phi_{q\,d}) \ ,
\end{equation}
which is a real $SU(4)$-invariant. This way of writing the determinant is very useful 
in the derivation the following identities, since it
heavily uses the completeness of $SU(4)$ indices, i.e.\ total anti-symmetrization in five/four
indices gives zero/the epsilon tensor. In the main text we introduced already 
the hermitian projection operators $\ca{P}$ and $\ca{F}$ and associated identities (\ref{projectors}),
(\ref{projectors2}). Other useful relations are:\\

\noindent
{\bf Basic relations}  
\begin{align}
  \label{mesa3}
  \Phi^{i\,a}\del_{a,j}|X|&=\delta^i{}_j|X|\ , \nonumber\\
  \del_{[i|a}|X|\,\del_{j]b}|X|&=\fr{|X|}{2}\,\del_{i\,a}\del_{j\,b}|X|\ , \nonumber\\
  \fr{1}{|X|}\, \del_{i\,a}\bar\del^{j\,b}|X|&=\ca{F}^a{}_c\,\del_{i\,a}\bar\del^{j\,c}|X|
    =\ca{F}^c{}_b\,\del_{i\,c}\bar\del^{j\,a}|X| \ ,
\end{align}
where $\del_{i\,a}:=\frac{\del}{\del\Phi^{i\,a}}$ and 
 $\bar\del^{i\,a}:=\frac{\del}{\del\bar\Phi_{i\,a}}=(\del_{i\,a})^\ast$. By complex conjugation of these
relations one obtains a similar set of identities. 
\\

\noindent
{\bf Super-derivatives}
\begin{align}
  \label{mesa2a}
  \nabla_{\alpha\,ij}\,\del_{k\,a}|X| &= \del_{k\,a}\nabla_{\alpha\,ij}|X|\nonumber\\
   \nabla_{\alpha\,ij}|X| & =-i\,(\Psi^c_{\alpha[i}\,\del_{j]c}|X|
       +\fr{1}{2}\varepsilon_{ijkl}\,\bar\Psi^k_{\alpha\,c}\,\bar\del^{l\,c}\bar\del^{k\,a}|X|)\nonumber\\
   \nabla_{\alpha\,ij}\ca{P}^a{}_b & = -\nabla_{\alpha\,ij}\ca{F}^a{}_b\nonumber\\
                                 & = \fr{i}{|X|}(\ca{P}^a{}_c\Psi^c_{\alpha[i}\,\del_{j]b}|X|
                         +\fr{1}{2}\,\varepsilon_{ijkl}\bar\Psi^k_{\alpha\,c}\ca{P}^c{}_b\bar\del^{l\,a}|X|).
\end{align}
These are basically the relations used to derive the expressions given in the main text.

\section{$SO(2,1)$ spinor conventions}\label{so3}

All spinors appearing in the main text, superspace coordinates or fields,  are Majorana spinors in $2+1$-dimensional 
space-time. Our  metric convention is
$\eta_{\mu\nu}=(-,+,+)$ and we choose a Majorana representation for the 
gamma-matrices\footnote{In terms of the Pauli matrices $\sigma^i$ for example 
$\gamma^0=-i\sigma^2\ ,\gamma^1=\sigma^1 , \gamma^2= \sigma^3$, see e.g.~\cite{RuizRuiz:1996mm} 
for more details.}
\begin{equation}
\label{eq:a1}
\{\gamma^\mu,\gamma^\nu\}^\alpha{}_{\beta}=2\eta^{\mu\nu}\delta^\alpha{}_{\beta}\ \ .
\end{equation}
Thus the matrices $\gamma^{\mu\ \alpha}{}_{\beta}$ are real and the Majorana condition on spinors imply 
that they are real two component spinors. Spinor indices are raised/lowered by the epsilon symbols with 
$\vep^{12}=\vep_{12}=1$ and choosing NW-SE conventions
\begin{equation}
\label{eq:a2}
\vep^{\alpha\gamma}\vep_{\beta\gamma}=\delta^\alpha_{\ \beta}\  ,\quad 
\lambda^\alpha:=\vep^{\alpha\beta}\lambda_\beta\Leftrightarrow\lambda_\beta 
    = \lambda^\alpha\vep_{\alpha\beta}\ .
\end{equation}
Introducing the real symmetric matrices 
$\sigma^\mu_{\alpha\beta}:=\gamma^{\mu\ \rho}{}_{\beta}\  \vep_{\rho\alpha}$ and 
$\bar{\sigma}^{\mu\ \alpha\beta}:= \vep^{\alpha\gamma} \vep^{\beta\delta} \sigma^\mu_{\gamma\delta}
=\vep^{\beta\rho}\  \gamma^{\mu\ \alpha}{}_{\rho} $ a three vector in spinor notation writes as 
a symmetric real matrix as 
\begin{equation}
\label{eq:a3}
 v_{\alpha\beta}:=\sigma^\mu_{\alpha\beta}\ v_\mu\ \ \Rightarrow \ \ 
  v^\mu=-\fr{1}{2}\bar{\sigma}^{\mu\ \alpha\beta}\ v_{\alpha\beta}\ \ ,\ \ 
   v^{\alpha\beta}w_{\alpha\beta}=-2\ v^\mu w_\mu\   .
\end{equation}
Another useful formula is
 \begin{equation}
   \vep^{\mu\nu\lambda}A_\mu B_\nu C_\lambda
         =\fr{1}{2}\ \vep^{\alpha\beta}A^{\gamma\delta}B_{\alpha\gamma}C_{\delta\beta}\ .
\end{equation}



\providecommand{\href}[2]{#2}\begingroup\raggedright\endgroup

\end{document}